\documentclass[preprint]{aastex}

\newcommand{\kms}{km~s$^{-1}$}
\newcommand{\ms}{m~s$^{-1}$}
\newcommand{\Wms}{W~m$^{-2}$}
\newcommand{\degrees}{$^{\circ}$}
\newcommand{\kvis}{\ensuremath{\kappa_{\mathrm{vis}}}}
\newcommand{\kir}{\ensuremath{\kappa_{\mathrm{IR}}}}
\newcommand{\Tint}{\ensuremath{T_{\mathrm{int}}}}
\newcommand{\Tirr}{\ensuremath{T_{\mathrm{irr}}}}

\begin{document}

\title{A General Circulation Model for Gaseous Exoplanets with Double-Gray Radiative Transfer}

\author{Emily Rauscher\footnote{NASA Sagan Fellow}
  \\ \textit{Lunar and Planetary Laboratory, University of Arizona,
  \\ 1629 East University Blvd., Tucson, AZ 85721-0092, USA}
  \\ and
  \\ Kristen Menou
  \\ \textit{Department of Astronomy, Columbia University,
  \\ 550 West 120th St., New York, NY 10027, USA}}

\begin{abstract}

We present a new version of our code for modeling the atmospheric circulation on gaseous exoplanets, now employing a ``double-gray" radiative transfer scheme, which self-consistently solves for fluxes and heating throughout the atmosphere, including the emerging (observable) infrared flux.  We separate the radiation into infrared and optical components, each with its own absorption coefficient, and solve standard two-stream radiative transfer equations.  We use a constant optical absorption coefficient, while the infrared coefficient can scale as a powerlaw with pressure.  Here we describe our new code in detail and demonstrate its utility by presenting a generic hot Jupiter model.  We discuss issues related to modeling the deepest pressures of the atmosphere and describe our use of the diffusion approximation for radiative fluxes at high optical depths.  In addition, we present new models using a simple form for magnetic drag on the atmosphere.  We calculate emitted thermal phase curves and find that our drag-free model has the brightest region of the atmosphere offset by $\sim$12\degrees~from the substellar point and a minimum flux that is 17\% of the maximum, while the model with the strongest magnetic drag has an offset of only $\sim$2\degrees~and a ratio of 13\%.  Finally, we calculate rates of numerical loss of kinetic energy at $\sim$15\% for every model except for our strong-drag model, where there is no measurable loss; we speculate that this is due to the much decreased wind speeds in that model.

\end{abstract}

\section{Introduction}

It has been almost one decade since the first atmospheric measurement of a hot Jupiter \citep{Charbonneau2002} and yet this class of exotic exoplanet still provides us with many mysteries waiting to be solved.  These include the culprit responsible for the stratospheric temperature inversions inferred for many hot Jupiters \citep{Hubeny2003,Fortney2008,Spiegel2009,Zahnle2009,Knutson2010}, the implication of a super-solar carbon-to-oxygen ratio in at least one hot Jupiter \citep{Madhu2011}, the possible influence of magnetic drag on the atmospheric circulation \citep{Perna2010a}, and a peculiar planet where dawn seems to be much hotter than noon \citep{Crossfield2010}.  In the midst of a growing collection of observations, some with unexpected or surprising results, is the continuing development of atmospheric models trying to interpret and understand the measurements.

One set of atmospheric models are numerical ones that simulate the global circulation patterns on close-in gas giants.  With masses and radii comparable to Jupiter, but subject to incident stellar fluxes 10,000 times stronger than what Jupiter receives from the Sun, and expected to be tidally locked into synchronous orbits, hot Jupiters exist in an atmospheric regime unlike anything in our solar system, and models of their atmospheric circulation are expanding into uncharted territory.  In order to try to understand how atmospheres work in this new regime, the models that have been developed so far represent a range of complexities, various approaches, and the use of different sets of assumptions.  Some of the simplest models solve the shallow water or equivalent barotropic equations and use various schemes to include the effect of radiative heating \citep{Cho2003,Cho2008,Langton2007}.  Others solve the primitive equations of meteorology, either using a Newtonian relaxation scheme for the radiative forcing \citep{Showman2002,Cooper2005,Showman2008,MR09,RM10}, a dual-band radiative transfer scheme \citep[][similar to the one we present here]{Heng2011}, or more complex non-gray radiative transfer \citep{Showman2009}.  Finally, there are also models that solve the full set of fluid equations, using dual-band flux-limited diffusion for the radiative transfer \citep{DobbsDixon2008,DobbsDixon2010}.

Here we present an updated version of our previous general circulation model that now includes a ``double-gray" radiative transfer scheme.  Fluxes throughout the atmosphere are separated into optical and infrared components, each with its own absorption coefficient, which is constant for the optical band and can scale as a powerlaw with pressure for the infrared band.  We use standard two-stream radiative transfer equations to solve for the vertical fluxes and calculate heating rates from those.  In addition to producing self-consistent radiative fluxes and heating rates, including the observable infrared flux emerging from the top boundary, this new code has the advantage of maintaining only a moderate level of complexity.  This facilitates comparison between simulated results and analytic profiles \citep[][see also Hansen 2008]{Guillot2010} and makes it easier to clearly identify the effects due to changes in opacity.

Our new code is described in detail in Section~\ref{sec:code}.  We demonstrate the functionality of our code by presenting a model of a generic hot Jupiter (Section~\ref{sec:models}) and we investigate issues related the atmosphere's behavior at deep pressures (Section~\ref{sec:deep}).  We then study the effect of magnetic drag on atmospheric circulation by applying a simplified drag scheme to our model, with drag strengths ranging from weak to strong (Section~\ref{sec:drag}).  In Section~\ref{sec:conc} we summarize our results.

\section{Description of the code} \label{sec:code}

The new version of our code presented here solves the primitive equations of meteorology\footnote{The primitive equations are a standard set of equations used to solve for the large-scale circulation of an atmosphere.  They are derived from the full fluid equations, solved in a rotating frame and subject to several simplifying assumptions \citep[see, e.g.,][]{Vallis}.} using the dynamical core originally developed by \citet{Hoskins1975}, converted for hot Jupiter studies in \citet{MR09} and used in subsequent papers \citep{RM10,Perna2010a,Perna2010b,Kempton2011}.  The previous version of this code (IGCM1) heated the atmosphere through a Newtonian relaxation scheme.  Here we use an updated version of the code \citep[IGCM3,][]{Forster2000} with several improvements, including a simple radiative transfer scheme, which we have adapted to model gaseous exoplanets.  This code also contains a standard dry convection scheme.  The static stability throughout each vertical column is calculated and any layers found to be convectively unstable are adjusted to constant potential temperature (entropy), with the vertical integral of enthalpy conserved.

\subsection{Radiative transfer} \label{sec:rad}

The primary driver of atmospheric circulation on close-in extrasolar planets is heating from the stellar irradiation, which is absorbed and re-emitted throughout the atmosphere.  We employ the common assumption that the incident radiation is predominantly at optical wavelengths, while the re-emitted radiation is in the infrared.  We can then separate our treatment of radiative transfer into two wavelength ranges, with downward optical flux absorbed as it penetrates deeper into the atmosphere and infrared flux absorbed and emitted isotropically throughout the atmosphere.  We calculate the upward and downward net fluxes in each vertical column of the atmosphere and then solve for the localized specific heating as: $dT/dt = (g/c_p)dF/dP$, where $dT/dt$ is the temperature tendency, $g$ is the gravitational acceleration (assumed to be constant through the model), $c_p$ is the specific heat, and $dF/dP$ is the vertical derivative of the net vertical flux (pressure, $P$, is our vertical coordinate).  Upward flux, in the direction of decreasing pressure, is defined to be positive.  For an atmosphere in local radiative equilibrium (i.e. without dynamics) $dF/dP=0$, either because there is a constant flux through the atmosphere (e.g. an outward heat flux from the planet interior) or because the optical heating is exactly balanced by infrared cooling.  The heating rate feeds into the energy equation of the dynamics solver ($Q$ in Equation 1d of Menou \& Rauscher 2009).

The incident optical flux at the top of each column of the atmosphere, with latitude $\phi$ and longitude $\theta$, is $F(\phi,\theta)=(1-A)F_{\mathrm{inc}} \cos(\phi) \cos(\theta)$ for the day side\footnote{The substellar point is at $\phi=\theta=0$ and the day side has $|\theta| \leq 90^\circ$.} and $F=0$ on the night side, where $F_{\mathrm{inc}}$ is the stellar flux incident on the top of the atmosphere at the substellar point and $A$ is the Bond albedo, which here we assume to be zero.  We assume a well-mixed optical absorber throughout the atmosphere, and negligible scattering, so that on the day side the downward flux in a column $\cos^{-1} \mu_\star$ degrees away from the substellar point ($\mu_\star = \cos(\phi) \cos(\theta)$) is given by the well-known equation \citep[e.g.,][]{Stephens1984}:
\begin{eqnarray}
F_{\downarrow \mathrm{vis}} (P) &=& (1-A)  \mu_\star F_{\mathrm{inc}} \exp\left(-\frac{1}{\mu_\star} \int_z^\infty \kvis du \right) \nonumber \\
	&=& (1-A)  \mu_\star F_{\mathrm{inc}} \exp\left(-\frac{1}{\mu_\star} \frac{\kvis}{g} P\right), \label{eqn:sw}
\end{eqnarray}  
\noindent where we have converted from path length ($du=\rho dz$) to pressure using the equation for vertical hydrostatic equilibrium (assumed in the primitive equations) and by integrating from the top of the atmosphere ($z=\infty$, $P=0$) down to the current pressure.  Note that the absorption of the incident flux, and the resulting heating of the atmosphere, is entirely determined by the wavelength-integrated absorption coefficient \kvis.  Lacking any reflection from a solid surface or clouds, the upward optical flux is always zero ($F_{\uparrow \mathrm{vis}}=0$).

Our assumption that all of the incoming stellar irradiation is at visible wavelengths leads to the upper boundary condition for our infrared flux: $F_{\downarrow \mathrm{IR}} (P=0) = 0$.  The bottom boundary condition is an upward heat flux from the interior: $F_{\uparrow \mathrm{IR}}(P=P_{\mathrm{0}})=\sigma_{\mathrm{SB}} \Tint^4$.  Throughout the atmosphere upward and downward infrared fluxes are absorbed and emitted as for a gray atmosphere, with a wavelength-integrated absorption coefficient that scales with pressure as a powerlaw:
\begin{equation}
\kir = \kappa_{\mathrm{IR},0} (P/P_{\mathrm{ref}})^{\alpha}. \label{eqn:kir}
\end{equation}
\noindent  The infrared fluxes are:
\begin{equation}
F_{\uparrow,\downarrow \mathrm{IR}}(P) = \int \left(1-\exp\left[-\frac{1.66}{g} \int \kir dP\right]\right) \frac{d \sigma T^4}{dP} dP \label{eqn:fir}
\end{equation}
\noindent where the integrations are from the bottom boundary ($P=P_0$) to the given pressure for the upward flux and from the top boundary ($P=0$) for the downward flux, and we have used the standard diffusivity factor 1.66 to account for integration of the isotropic radiation over all zenith angles \citep[see, e.g.,][]{Stephens1984}.

If both the infrared and optical absorption coefficients are constant with pressure ($\alpha=0$) our radiative scheme (with the dynamics turned off) should reproduce the analytic profiles from \citet[][see also Hansen 2008]{Guillot2010}.  Note that for this type of ``double-gray" atmosphere with constant absorption coefficients, the temperature-pressure profiles can never be convective; as shown in Appendix~\ref{sec:conv}, for a convective region to exist the powerlaw index, $\alpha$, must be greater than a threshold value based on the ratio of the specific gas constant to the specific heat ($R/c_p$).

We initialize our model runs with no winds and with temperatures everywhere set to the analytic night side profile (Equation 27 of Guillot 2010).  For the first (planet) day of the run, the fluxes from the radiative transfer scheme are not used and instead a Newtonian relaxation scheme heats each point in the atmosphere toward the value it would have in local radiative equilibrium, with a relaxation timescale set to 0.1 planet days.  During this time the atmosphere is also allowed to respond by developing winds.  After the first day the Newtonian scheme is no longer used and the heating is entirely determined by fluxes from the radiative transfer scheme.

To test the implementation of our radiative transfer in the code, we set the infrared opacity to be constant with pressure ($\alpha=0$ in Equation~\ref{eqn:kir}) and compared our temperature-pressure profiles against the analytic solutions from \citet{Guillot2010}.\footnote{Analytic profiles are also calculable for non-constant absorption coefficients, as shown in Appendix~\ref{sec:tpprofiles}; see also \citet{Heng2011}.  When testing our radiative transfer scheme, we also checked how our code performed for non-constant absorption coefficients and found our results to be in good agreement with the analytic profiles, after the modification described in Section~\ref{sec:fdiff}.}  We ran several tests in which the downward optical flux at the top boundary was identical at all points around the globe, so that there are no horizontal gradients and no atmospheric motion.  This effectively turns the code into a one-dimensional radiative transfer solver.  For a given set of absorption coefficients and boundary fluxes, we began with initial profiles set to the analytic solutions and then let the code run until each column was in local radiative equilibrium: $dF/dP = 0$, with heating equal to cooling, or with a constant flux set by the bottom boundary condition: $F_{\uparrow,\mathrm{IR}}=\sigma_{\mathrm{SB}} \Tint^4$.  We considered a profile to be equilibrated once it was within less than 1\% of $dF/dP=0$.

\begin{figure}[ht!]
\begin{center}
\includegraphics[width=0.475\textwidth]{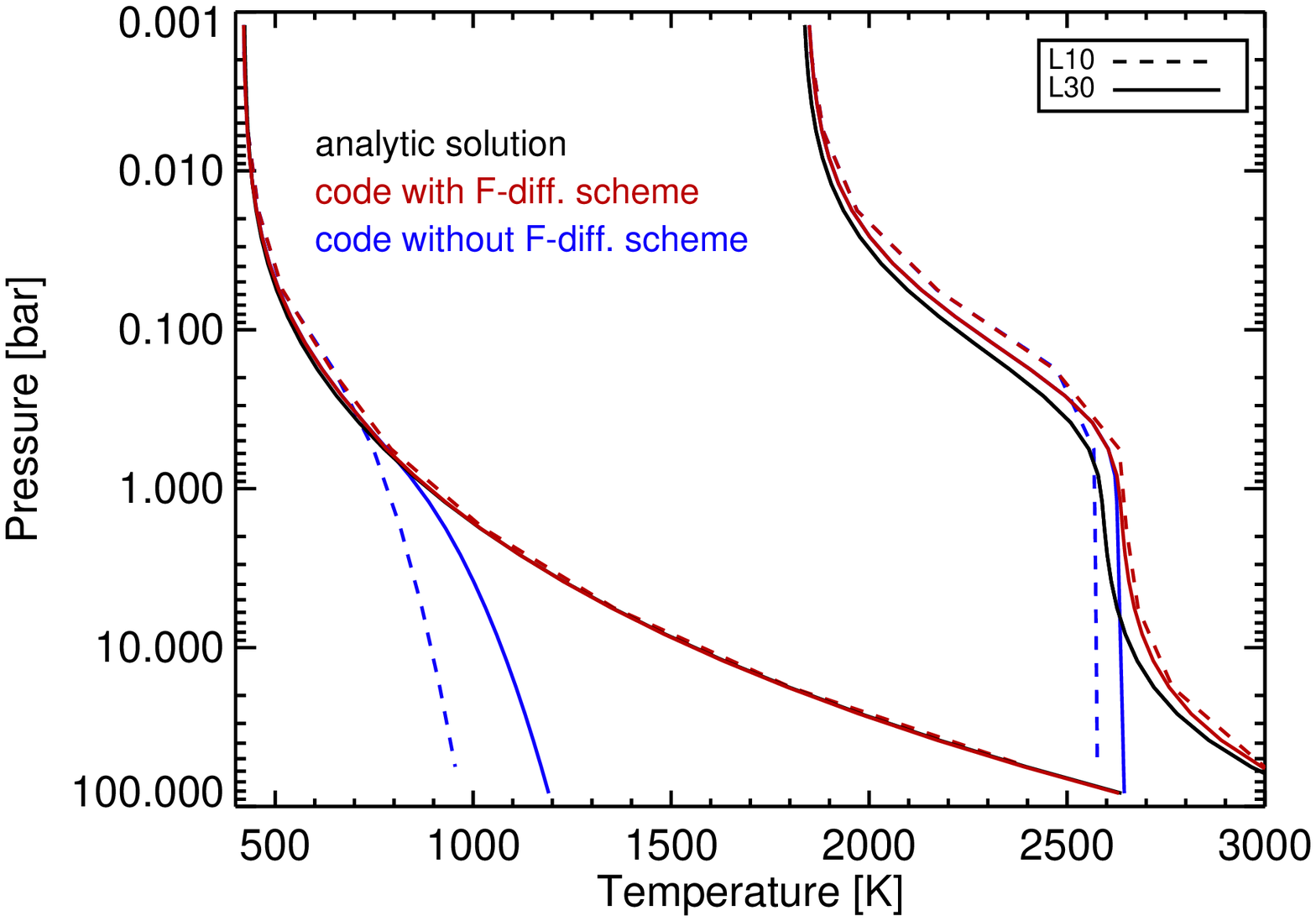}
\includegraphics[width=0.475\textwidth]{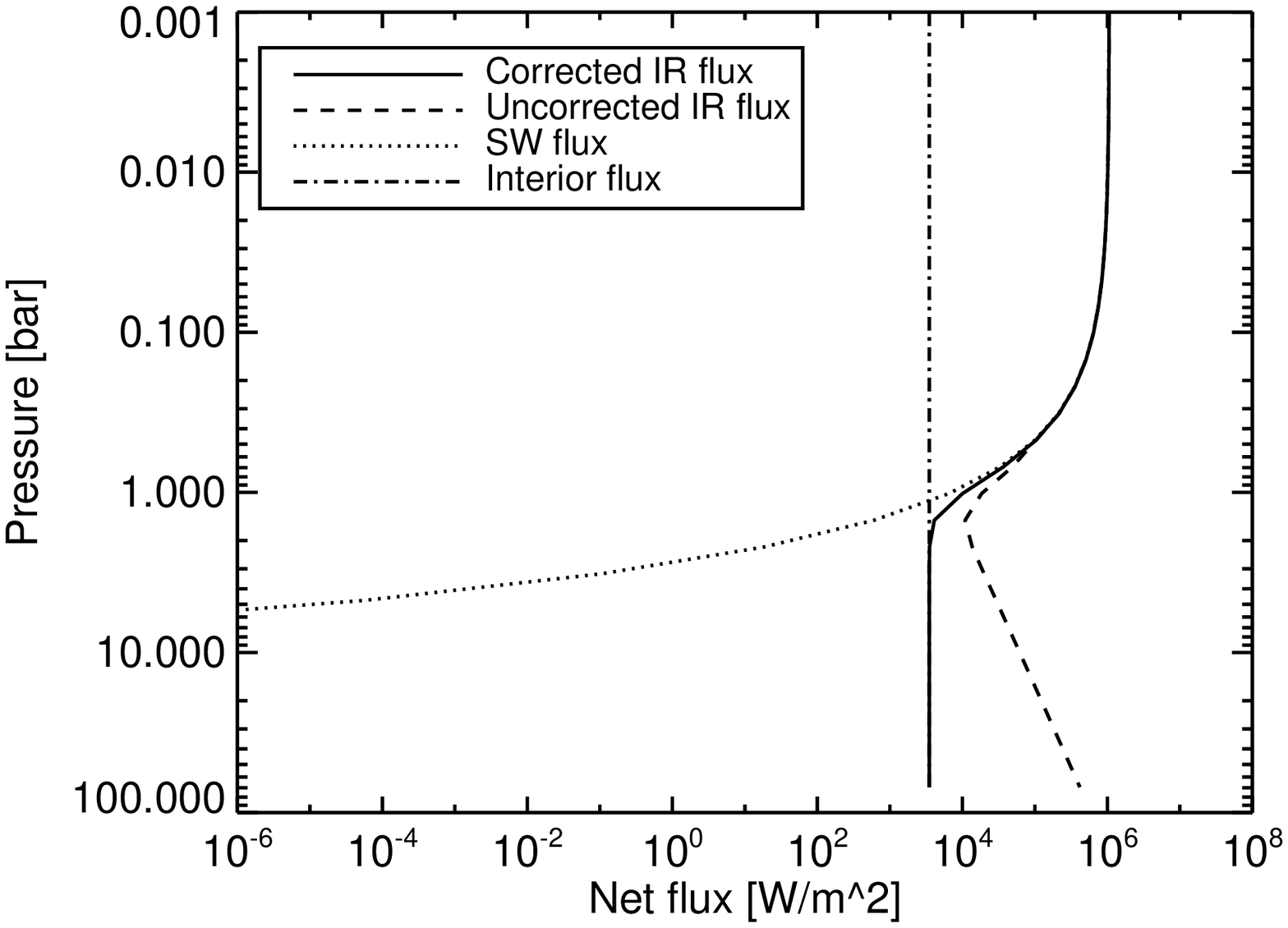}
\end{center}
\caption{\emph{Left:}  temperature-pressure profiles produced by our radiative transfer code, both with and without a transition to using the diffusion approximation at high optical depth.  Shown are the substellar and night side profiles, for models with 10 or 30 vertical levels from 1 mbar to 100 bar.  Plotted for comparison are the analytic solutions (for the same choice of parameters) from \citet{Guillot2010}.  Our flux-limited diffusion scheme produces results that agree very well with the analytic profiles.  \emph{Right:} a profile of the vertical fluxes at the substellar point.  High in the atmosphere (at low pressures) the downward optical (``shortwave," SW) flux is balanced by the upward infrared (IR) flux.  Below the optical photosphere the infrared flux (``corrected," calculated using the flux-limited diffusion scheme) matches the upward flux from the interior, chosen as our bottom boundary condition.  Also plotted is the ``uncorrected" infrared flux, calculated from the radiative transfer without the flux-diffusion scheme.  This extra upward flux would have cooled the deep layers of the atmosphere, as evidenced by the uncorrected profiles in the lefthand plot.} \label{fig:fdiff}
\end{figure}

This testing showed that, although our profiles matched the analytic solution at low pressures, they tended toward isothermal in the deepest levels, rather than approaching an adiabat (as shown in Figure~\ref{fig:fdiff}).  The reason for this is that our numerical domain spans many orders of magnitude in pressure and vertical levels are evenly distributed in $\log P$.  Deep in the atmosphere the pressure difference between adjacent levels becomes very large ($dP \sim P$), the optical depth greatly exceeds unity ($\tau=\kir P/g$), and the exponential term in Equation~\ref{eqn:fir} goes to zero.  When the deepest levels are in radiative equilibrium, the temperature profile is entirely determined by the vertical resolution, with the temperature differences between adjacent pressure levels set by: $\Delta (\sigma T^4) = \sigma \Tint^4$ (a profile that exactly matches those shown in Figure~\ref{fig:fdiff}).

\subsection{Transition to flux-limited diffusion at high optical depth} \label{sec:fdiff}

In order to achieve the correct temperature-pressure profiles at depth, we have implemented a scheme in which we transition to calculating the radiative flux using the diffusion approximation at optical depths greater than one.  The flux through the optically thick, deep atmosphere is calculated as:
\begin{equation}
F_{\mathrm{IR,diff}} = \frac{16 g \sigma T^3}{3 \kir} \frac{dT}{dP}. \label{eqn:fdiff}
\end{equation}
\noindent A temperature profile that increases with pressure results in an upward flux ($F_{\mathrm{IR}} > 0$).

The transition from fluxes calculated using the main radiative transfer equations (\ref{eqn:sw}, \ref{eqn:fir}) and the diffusion approximation (\ref{eqn:fdiff}) occurs based on the local optical depth ($\tau = \kir P/g$, taking into account any dependence of $\kir$ with pressure).  The original radiative scheme begins to fail when the term $E=\exp(-[1.66/g]\int \kir dP)$ becomes too small, and so we use this as a criterion for when to begin the implementation of the flux diffusion scheme.  Through trial and comparison to the analytic solutions, we determined that the fix should begin at the point where $E \leq 0.01$, which in our code corresponds to an optical depth $\tau > \tau_{\mathrm{limit}} = 5.55/(10^A-10^{-A})$, where $A$ is the vertical resolution of the model, $A=\log_{10}(P_{max}/P_{min})/NL$.  For a model whose bottom boundary is at $P_{max}=100$ bar, whose upper boundary is $P_{min}=1$ mbar, and that has $NL=30$ levels, $\tau_{\mathrm{limit}}=7$ ($= 0.56$ bar in Figure~\ref{fig:fdiff}).  A model using only 10 vertical levels would have $\tau_{\mathrm{limit}}=2$ ($= 0.16$ bar in Figure~\ref{fig:fdiff}).  This resolution-dependent $\tau_{\mathrm{limit}}$ appropriately fixes the resolution dependence of the problem with the original radiative transfer scheme.

At optical depths greater than $\tau_{\mathrm{limit}}$, the infrared flux is calculated as: $F_{\mathrm{IR}}(P) = (E^{0.023}) F_{\mathrm{IR,rad}} + (1-E^{0.023}) F_{\mathrm{IR,diff}}$, where $F_{\mathrm{IR,rad}}$ is the infrared flux calculated from Equation~\ref{eqn:fir}.  The transition to the diffusion fluxes has the same exponential dependence as the problem with the original radiative scheme and the factor of 0.023 was chosen so that at the pressure where $E=0.01$ ($\tau=\tau_{\mathrm{limit}}$), $F_{\mathrm{IR}} = 0.9 \times F_{\mathrm{IR,rad}} + 0.1\times F_{\mathrm{IR,diff}}$.  We found that this correction produced a smooth transition (both in fluxes and the temperature-pressure profiles) from the optically thin to optically thick regimes.

\subsection{Advantages of this scheme}

There are several advantages to using this radiative transfer scheme.  The main benefit over the Newtonian relaxation scheme is that the atmosphere is heated through self-consistently calculated fluxes, given a small set of parameters: the optical absorption coefficient ($\kvis$), the infrared absorption constant ($\kappa_{\mathrm{IR},0}$) and powerlaw index ($\alpha$), the external stellar flux ($F_{\mathrm{inc}}$), and the internal heat flux ($F_{\mathrm{int}}$).  The upward flux at the top boundary of the model can be used to create maps of the infrared flux emitted by the planet.  Although detailed physics are hidden within the two absorption coefficients, the relative simplicity of this radiative transfer facilitates comparison to analytic predictions.  In particular, the dynamic temperature-pressure profiles can easily be compared to analytic profiles for local radiative equilibrium \citep[see][and Appendix~\ref{sec:tpprofiles} for profiles with non-constant infrared absorption coefficients]{Guillot2010}.

\section{A fiducial hot Jupiter atmospheric model} \label{sec:models}

We present a fiducial model for a hot Jupiter with global parameters similar to HD 209458b, as listed in Table~\ref{tab:params}.  We chose HD 209458b parameters so that we could use absorption coefficients, internal, and external heat fluxes based on values from \citet{Guillot2010} and, in Section~\ref{sec:drag}, magnetic drag timescales from \citet{Perna2010a}.  Although there is evidence for a temperature inversion in the atmosphere of HD 209458b \citep{Knutson2008}, we do not include stratospheric absorbers in our model, in the interest of generality.

\begin{deluxetable}{lcccc}
\tablewidth{0pt}
\tablecaption{Parameters for our fiducial model}
\tablehead{
\colhead{Parameter}  &  \colhead{Value} & \colhead{Units}
}
\startdata
Radius of the planet, $R_p$ & $1\times 10^8$ & m \\
Rotation rate, $\Omega$ & $2.1 \times 10^{-5}$ & s$^{-1}$ \\
Gravitational acceleration, $g$ & 8 & m s$^{-2}$ \\
Specific gas constant, $\mathcal{R}$ & 3523 & J kg$^{-1}$ K$^{-1}$ \\
Ratio of gas constant to heat capacity, $\mathcal{R}/c_P$ & 0.286 & -- \\
Optical absorption coefficient, \kvis & $4 \times 10^{-3}$ & cm$^2$ g$^{-1}$ \\
Infrared absorption coefficient, $\kappa_{\mathrm{IR},0}$  & $1 \times 10^{-2}$  & cm$^2$ g$^{-1}$ \\
Infrared absorption powerlaw index, $\alpha$ & 0 & -- \\
Infrared absorption reference pressure, $P_{\mathrm{ref}}$ & 1 & bar \\
Internal heat flux, $F_{\uparrow \mathrm{IR}, \mathrm{int}}$ & 3500 & W m$^{-2}$ \\
\ \ \ Corresponding temperature, \Tint & 500 & K \\
Incident flux at substellar point, $F_{\downarrow \mathrm{vis}, \mathrm{irr}}$ & $1.06 \times 10^6$  & W m$^{-2}$  \\
\ \ \ Corresponding temperature, \Tirr & 2078 & K \\
\enddata
\label{tab:params}
\end{deluxetable}
\clearpage

We used a horizontal spectral resolution of T31, which corresponds to $\sim$4\degrees~in latitude and longitude, and used 30 vertical levels, evenly spaced in pressure from 1 mbar to 100 bar.  We ran the model for 2000 planet days ($= 2000$ orbital periods), by which point the kinetic energy and average temperature of the atmosphere have stabilized, after the initial spin-up period.  All of the results shown in this paper, both for this fiducial model and those in later sections, are for a snapshot of the atmosphere at day 2000.  In terms of variability, there is a very minor level of quantitative change from day to day, and on large scales the atmospheric structure is very steady.

We performed partial runs at T21 and T42 and found consistency with the T31 results shown here, but the T42 resolution was too computationally expensive for this work.  We also tested the strength and order of the hyperdissipation we used to prevent the build up of noise on small scales, which takes the form: $dX/dt=\nu_{\mathrm{diss}} (-1)^{p+1}\nabla^{2p}X$, where $X$ represents the vorticity, divergence, and temperature fields.  For horizontal resolutions of T21, T31, and T42 we tested orders of $p=2$, 3, and 4, and values of $\nu_{\mathrm{diss}}$ corresponding to e-folding times (for damping the smallest scales) at fractions of a planet day: 0.5, 0.05, 0.005, and 0.0005.  The flow quickly became numerically unstable for a damping time of 0.5 and exhibited an obvious build-up of kinetic energy at the smallest scales for 0.05.  A comparison of the kinetic energy spectra revealed that the $p=2$ order was overdamping kinetic energy on all scales.  We chose to use an order of $p=4$ and a damping timescale of 0.005 planet days for our T31 fiducial model.  Since the radiative timescales throughout our model vary by orders of magnitude, the upper levels may be underdamped and the deepest levels overdamped \citep{Thrastarson2011}, but we use this form for hyperdissipation until a more robust and physically motivated scheme is developed.

\begin{figure}[ht!]
\begin{center}
\includegraphics[width=0.6\textwidth]{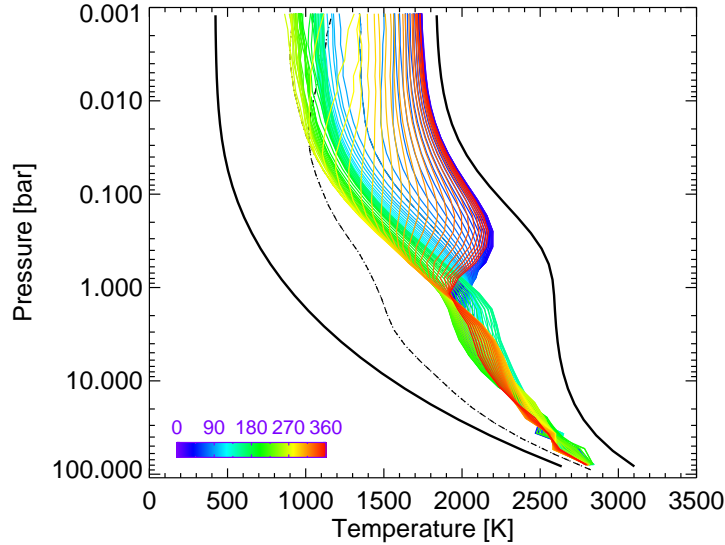}
\end{center}
\caption{Temperature-pressure profiles for the fiducial run.  The dashed and dotted lines (that plot on top of each other) give the profiles for the north and south poles.  All other lines are for atmospheric columns along the equator, where the color indicates the longitude.  The thick black lines give the analytic profiles for the substellar point and night side, if they were in local radiative equilibrium.  High in the atmosphere (at low pressures) the main temperature difference is between day and night, while at moderate pressures the equator-pole difference dominates.  At the deepest pressures the temperatures are well homogenized.} \label{fig:d_tprof}
\end{figure}

Figure~\ref{fig:d_tprof} shows temperature-pressure profiles throughout the atmosphere for our fiducial model.  Also plotted are the analytic, local radiative equilibrium profiles for the substellar point and the night side (previously seen in Figure~\ref{fig:fdiff}).  As is typical of previous hot Jupiter models, there is a strong day-night temperature difference at high altitude, which decreases with increasing pressure.  At moderate pressures ($\sim$1 bar) the primary temperature difference is between the equator and poles.  There is also a dynamically created temperature inversion at these pressures.  At low pressures the dayside remains at temperatures close to radiative equilibrium, but at deeper pressures the winds are better able to alter the temperature structure and the day side is cooled from more efficient homogenization with the night side.  Figure~\ref{fig:d_tprof} also shows that our use of the flux-limited diffusion scheme produces the correct behavior in the pressure-temperature profiles at depth.

\begin{figure}[ht!]
\begin{center}
\includegraphics[width=0.5\textwidth]{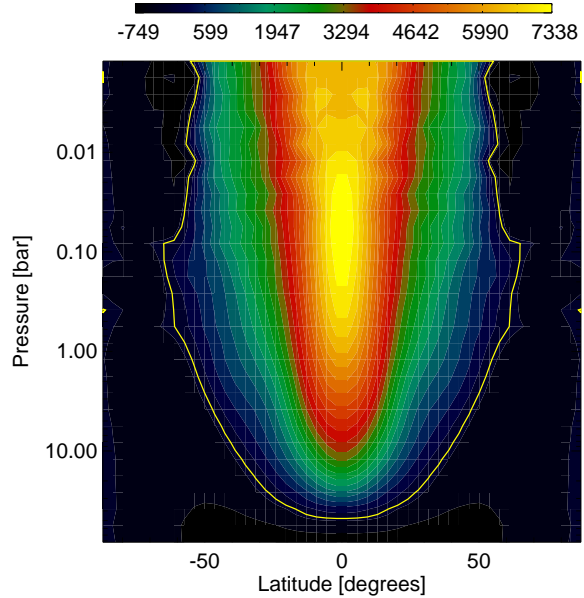}
\end{center}
\caption{The zonal average of the zonal (east-west) wind in \ms, as a function of latitude and pressure, for the fiducial run.  The yellow line separates eastward (positive) flow from westward (negative) flow.  An equatorial super-rotating jet extends throughout most of the atmosphere, typical of hot Jupiter circulation.} \label{fig:d_uz}
\end{figure}

Figure~\ref{fig:d_uz} shows the zonal average of the zonal wind for our fiducial model, with the equatorial super-rotating jet that is standard in hot Jupiter models \citep{Showman2011}.  This jet extends from the uppermost levels to almost the bottom of the atmosphere, with only the two deepest levels showing westward equatorial flow.  Whereas at high altitudes the equatorial jet is accompanied by strong flow (zonal and meridional) from day to night across most regions of the terminator, the equatorial jet dominates the flow for levels below the optical photosphere.  Also in agreement with most hot Jupiter models, we find strongly supersonic winds high in the atmosphere (at several times the sound speed), a transonic flow at moderate pressures, and subsonic (only) flow at pressures greater than $\sim$10 bar.

\begin{figure}[ht!]
\begin{center}
\includegraphics[width=0.95\textwidth]{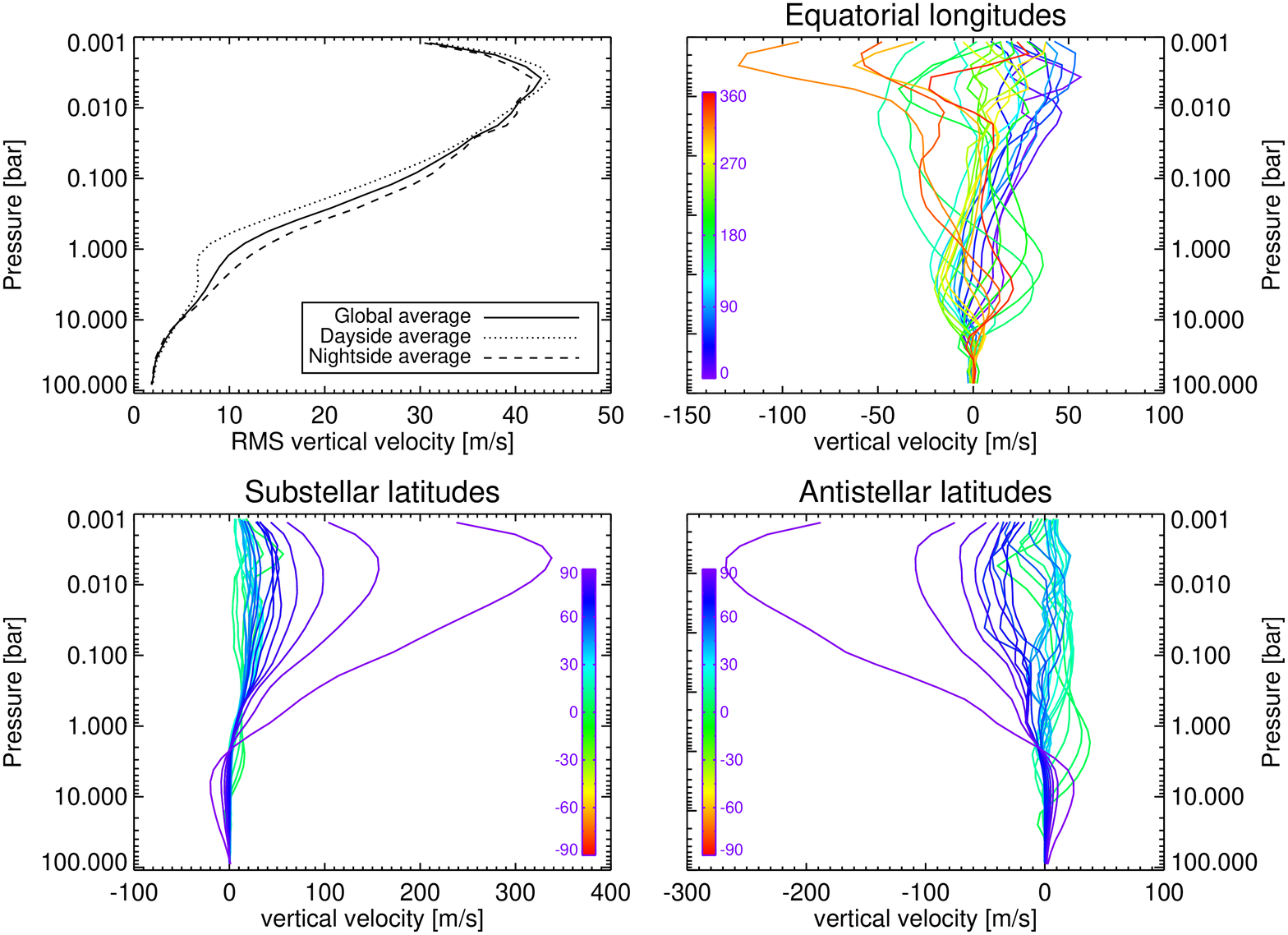}
\end{center}
\caption{Vertical velocities in our fiducial model.  In all cases the trend toward zero velocity at low pressures is set by the top boundary condition.  \emph{Upper left:} the root-mean-squared velocity as a function of pressure, horizontally averaged over the day side, night side, or entire globe.  \emph{Upper right:} vertical velocities for individual atmospheric columns at the equator, for the full range of longitudes (in color).  \emph{Bottom:} vertical velocites for columns at a longitude of 0\degrees~(substellar, \emph{left}) or 180\degrees~(antistellar, \emph{right}), for the full range of latitudes (in color).  Strong hemispheric symmetry means that the northern profiles (latitude $>0$\degrees) are plotted over the southern profiles.  While in general vertical velocities are low, in localized regions they can exceed 100 \ms.} \label{fig:d_vert}
\end{figure}

In Figure~\ref{fig:d_vert} we plot vertical velocities throughout the model.  Although the primitive equations used in this code replace the vertical momentum equation with hydrostatic balance, vertical motions are still present through the continuity equation: $d\omega/dP = -\nabla \cdot \vec{v}$, where $\omega = dP/dt$ and $\vec{v}$ is the horizontal wind.  We integrate the continuity equation down from the top of the model, where the boundary condition imposes $\omega=0$ at $P=0$.  We then convert to a vertical velocity ($dz/dt$) through use of the hydrostatic equation: $dP/dz = - \rho g$.  We note that snapshots of the vertical wind profiles taken at other points in the run match the main features shown here.

The strength of vertical motion in the atmosphere has important implications for the mixing of chemical species throughout the atmosphere, which can introduce chemical disequilibrium and influence the observable properties of the planet \citep{Moses2011}.  For example, titanium oxide (TiO) and vanadium oxide (VO) have been identified as possible candidates for the absorbing species responsible for stratospheric temperature inversions, but \citet{Spiegel2009} demonstrated that very strong vertical mixing would be required to keep these absorbers aloft in the atmosphere.  From Figure~\ref{fig:d_vert} we can see that vertical speeds in our model are generally very slow, at tens of meters per second (compared to horizontal wind speeds of $\sim$1 \kms), although in localized regions they can exceed 100 \ms.  Throughout most of the atmosphere the vertical winds are systematically stronger on the night side than the day side, although at pressures less than $\sim10$ mbar the day side winds are slightly stronger.  There is a trend toward higher vertical speeds at lower pressures and the polar regions have the strongest vertical motion, upward on the day side and downward on the night side.  The direction of vertical motion is consistent with the substellar-to-antistellar flow at low pressures; the divergent flow on the day side causes upward vertical motion, while convergent flow on the night side results in downward motion.  The profiles in Figure~\ref{fig:d_vert} demonstrate that the strength of vertical mixing depends on the local atmospheric flow structure.  However, the horizontal wind speeds are much stronger than vertical wind speeds, so that horizontal mixing can also bring species out of equilibrium \citep{Cooper2006} and can add to the difficulty of keeping TiO/VO aloft \citep{Showman2009}.  The amount of chemical mixing throughout the atmosphere is inherently a three-dimensional problem.

\begin{figure}[ht!]
\begin{center}
\includegraphics[width=0.45\textwidth]{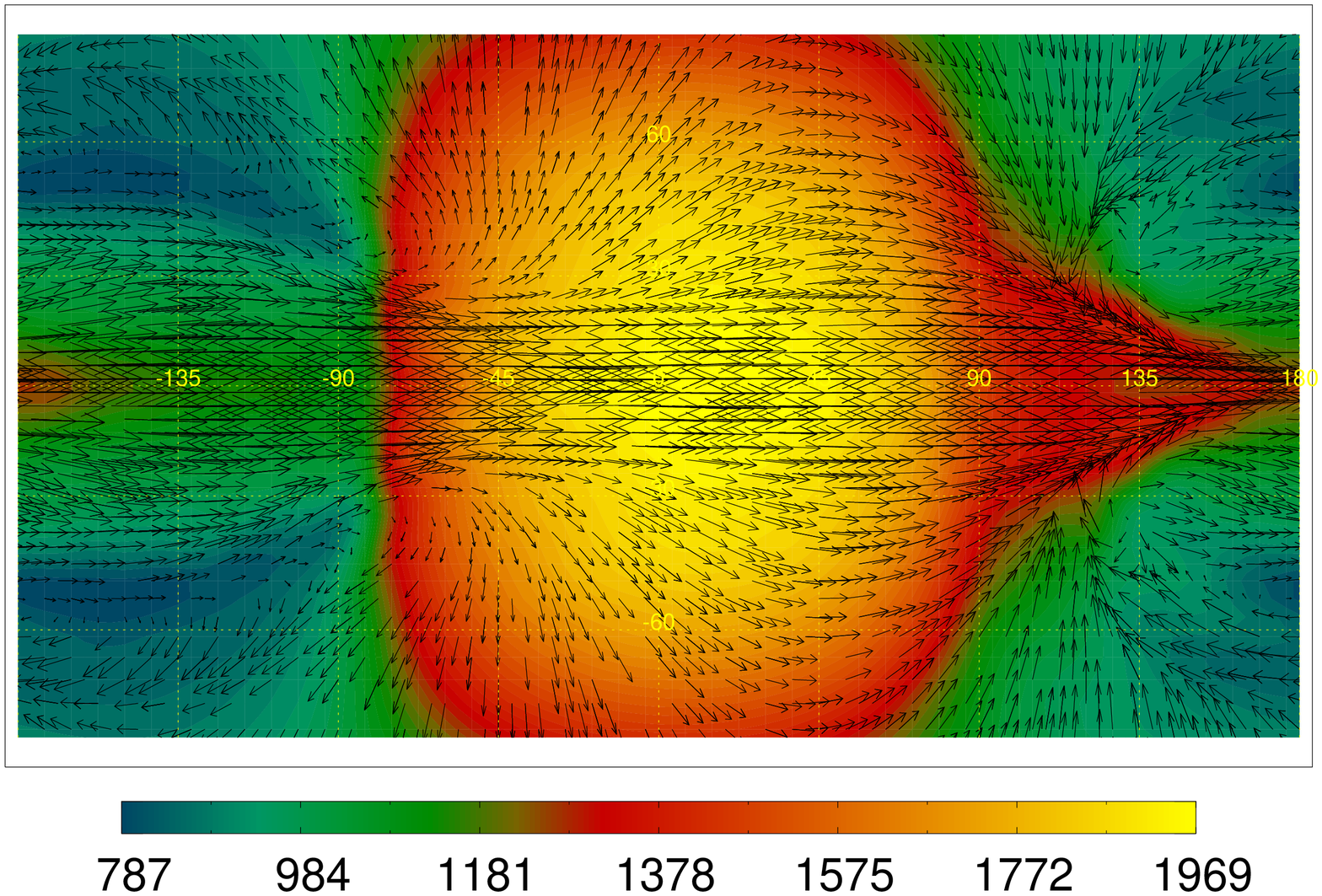}
\includegraphics[width=0.45\textwidth]{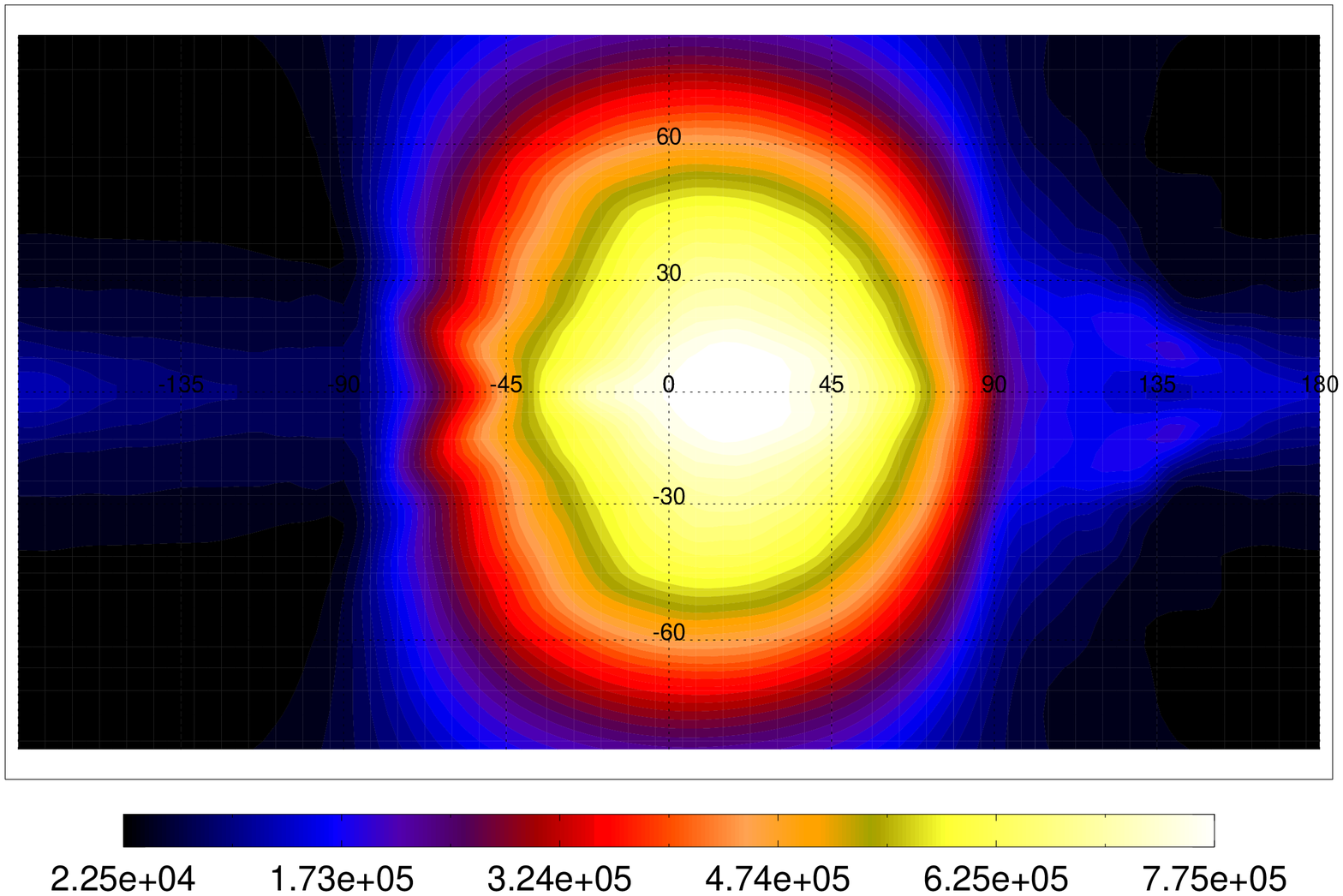}
\end{center}
\caption{\emph{Left:} The infrared photosphere of the planet for our fiducial model, shown in cylindrical projection with the substellar point at the center of the plot.  The color gives the temperature [in K] and the arrows show the wind vectors.  The maximum wind speed at this level is 8 \kms.  \emph{Right:} A cylindrical map of the infrared flux emitted from the top boundary of the model [in \Wms].  Note that the hottest region of the photosphere is advected eastward of the substellar point and this results in the emitted flux having a maximum at a longitude $\approx15$\degrees.}  \label{fig:d_phot}
\end{figure}

Figure~\ref{fig:d_phot} shows the temperature and flow pattern at the infrared photosphere ($\tau = 2/3$, $P=53$ mbar), as well as a map of the infrared flux emitted from the top boundary of the model.  We again see typical hot Jupiter atmospheric structure, where there is a significant day-night temperature difference, but the equatorial jet has advected the hottest region of the atmosphere eastward of the substellar point.  Our radiative scheme self-consistently produces the infrared map that would be observed for this planet, from the upward infrared flux at the top boundary of the model.

\begin{figure}[ht!]
\begin{center}
\includegraphics[width=0.47\textwidth]{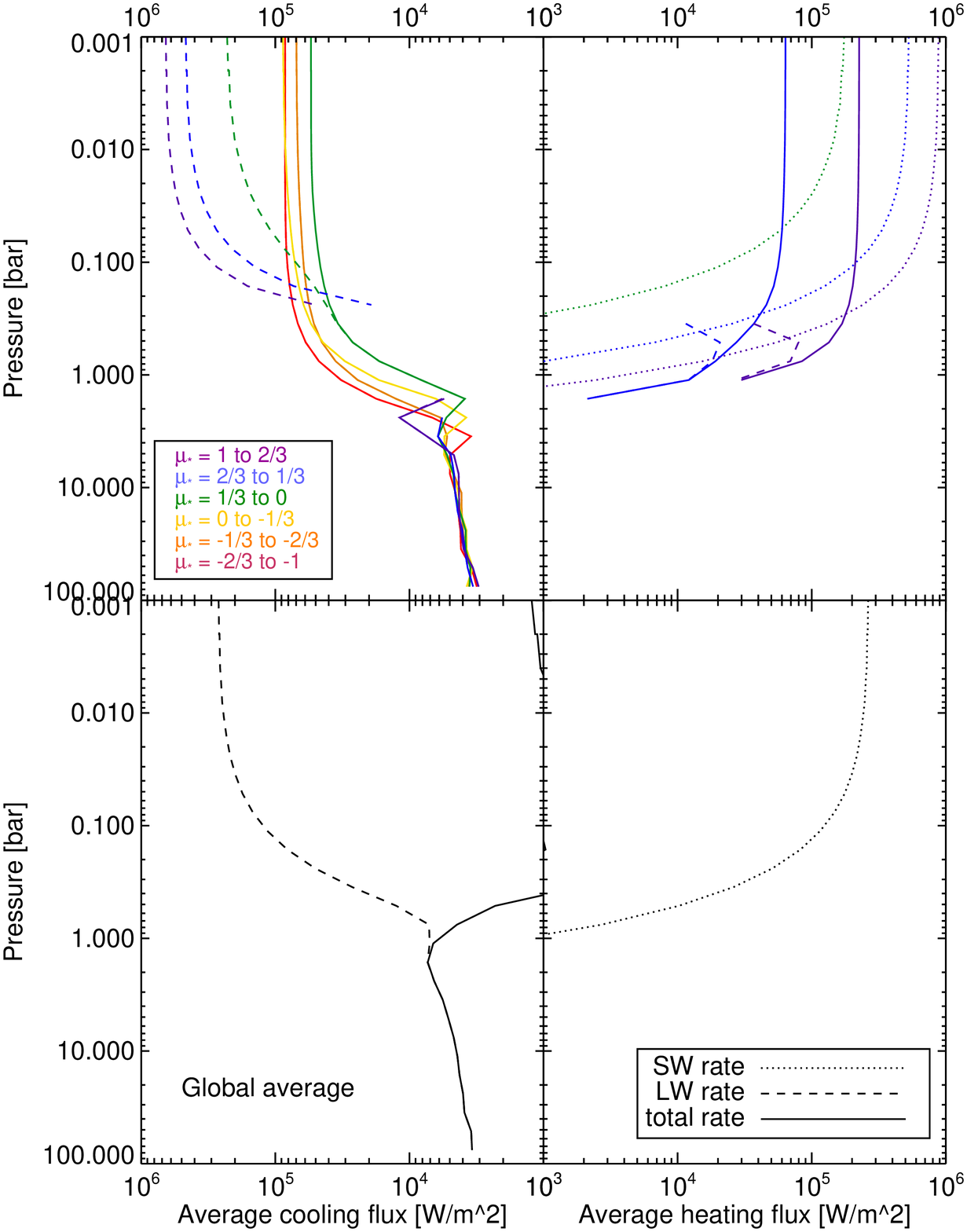}
\includegraphics[width=0.47\textwidth]{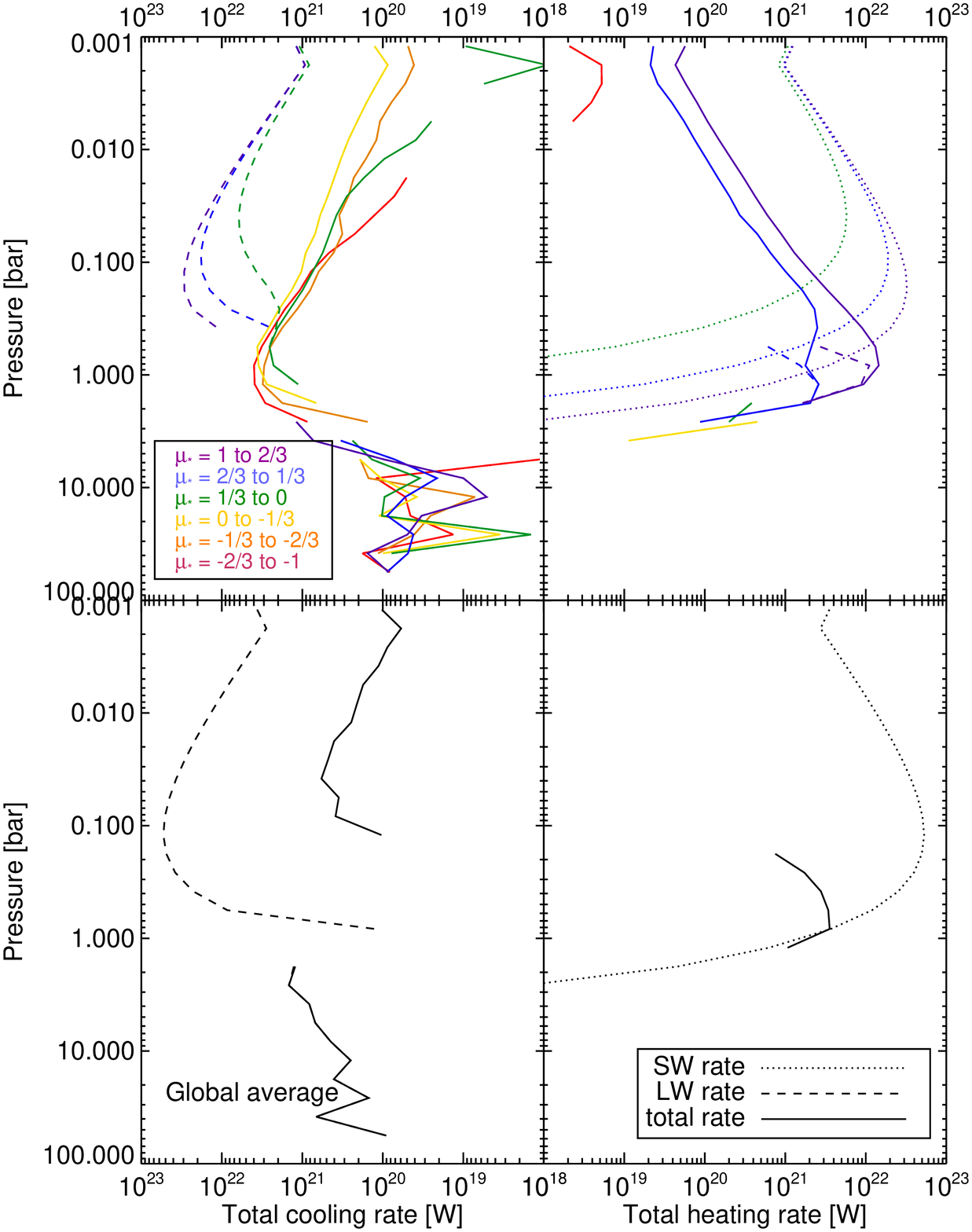}
\end{center}
\caption{Self-consistent fluxes and heating rates throughout the model at planet day 2000.  The left set of plots shows the cooling (upward) and heating (downward) fluxes throughout the atmosphere, while the right set gives the resulting heating and cooling rates.  The optical (``shortwave," SW) rates are plotted as dotted lines, the infrared (``longwave," LW) rates are plotted as dashed lines, and the total values are plotted as solid lines.  The colored lines show rates that have been averaged over equal-area annuli of the atmosphere, in increasing angle from the substellar point (purple) to the anti-stellar point (red), where $\mu_\star$ is the cosine of the angle from the substellar point.  The black lines show the horizontally averaged rate at each level.  Gaps in the profiles indicate regions where the flux or rate has switched sign, between heating and cooling; for clarity we do not plot single layers with sign fluctuations.} \label{fig:d_heating}
\end{figure}

One of the advantages of this new code is that the radiative scheme self-consistently solves for the infrared and optical fluxes throughout the atmosphere and calculates the resulting local diabatic heating throughout the three-dimensional atmosphere.  In Figure~\ref{fig:d_heating} we show the fluxes and heating for various regions of the atmosphere.  The optical flux is only incident on the day side, is always downward, and heats the atmosphere, while the infrared flux is in both directions throughout the planet, but is primarily upward and cools the atmosphere.  

For an atmosphere in local radiative equilibrium, at high pressures (below the optical photosphere) there would be a non-zero cooling flux, set by the boundary condition of upward flux from the planet interior (see Figure~\ref{fig:fdiff}).  The incident stellar flux sets the top boundary condition and high in the atmosphere the optical heating flux is mostly balanced by the infrared cooling flux, with a slight offset due to the cooling flux coming from the interior.  The heating at each level in the atmosphere is the difference between the net fluxes into and out of the layer and, by definition, would be near zero everywhere for local radiative equilibrium (modulo the small interior heat flux leaking out through the atmosphere).

The left panel of Figure~\ref{fig:d_heating} shows that the interplay between radiation and atmospheric dynamics results in flux profiles altered from what we would expect for local radiative equilibrium.  At the bottom of the atmosphere the cooling flux is close to the upward flux of 3500 \Wms~from the planet interior (set as a boundary condition).  In the upper atmosphere the infrared and optical fluxes are close to being globally balanced.  The top two plots show that the optical flux only passes through the day side of the planet ($\mu_\star = \cos(\phi) \cos(\theta) > 0$) and that the regions closest to the substellar point (purple line) experience a stronger optical flux that penetrates deeper than areas close to the terminator (green line).  The flux at the top of the model is set by the incident stellar flux boundary condition, here chosen to be $1.06 \times 10^6$ \Wms~at the substellar point (and decreasing as $\mu_\star$, according to Equation~\ref{eqn:sw}).

There is a net heating flux throughout most of the day side (aside from the region near the terminator), with the infrared cooling fluxes unable to completely balance the optical heating.  The total dayside heating flux extends deeper than the optical photosphere because of the contribution from infrared flux for a few layers around 1 bar.  This is due to the dynamically-induced thermal inversion at these pressures (see Figure~\ref{fig:d_tprof}), which leads to the net downward transport of infrared radiation.

The right panel of Figure~\ref{fig:d_heating} shows horizontally integrated diabatic heating rates throughout the atmosphere, which are calculated from the fluxes shown in the left panel; locally $dT/dt=(g/c_p)dF/dP$, where upward flux (in the direction of lower pressure) is defined to be positive (see Section~\ref{sec:rad}).  When flux increases with increasing pressure, there is heating; this is the case for the downward (negative) optical flux at the substellar point, which goes to zero as pressure increases.  Likewise, in the upper atmosphere the upward (positive) infrared flux near the antistellar point decreases with decreasing pressure, which heats this region of the night side.  For the globally averaged rates, the optical flux always leads to heating while the infrared generally leads to cooling.  The net globally averaged profile has cooling throughout most of the atmosphere, aside from a region near the optical photosphere where there is net heating.

The global net heating rate is $3.9\times 10^{21}$ W.  A planet in global radiative equilibrium should have a net radiative heating rate of zero.  However, for a steady state atmosphere, the dissipation of wind kinetic energy (numerical or physical) must be balanced by a non-zero conversion from thermodynamic potential energy, which in turn must be generated by a non-zero heating rate \citep[e.g.][]{Pearce1978}.  Physically, the dissipation of kinetic energy leads to frictional heating and the energetics are balanced.  However, in our fiducial model the numerical dissipation of kinetic energy (through hyperdissipation or other numerical loss) is not returned to the atmosphere as heating and so the net radiative heating must be non-zero to compensate.  We can take the ratio of the net heating rate and the rate of energy input into the model (the total of the incident stellar flux and the cooling flux from the interior) to determine that in this model the rate of numerical loss of kinetic energy is 12\%, a significant value and similar to the numerical loss found in our previous version of this code \citep{RM2011}.

\subsection{The effects of internal heating and flux diffusion on the deep atmosphere} \label{sec:deep}

The temperature structure of the upper atmosphere, above the optical and infrared photospheres, is dominated by the external stellar heating.  At deeper pressures, however, the atmospheric structure is set by the strength of the heat flux from the interior or, equivalently, the entropy of the interior adiabat.  Here we present two additional models to examine how the behavior of the deepest levels of our model depends on our choice of \Tint~and our use of the flux diffusion scheme (Section~\ref{sec:fdiff}).

\begin{figure}[ht!]
\begin{center}
\includegraphics[width=0.475\textwidth]{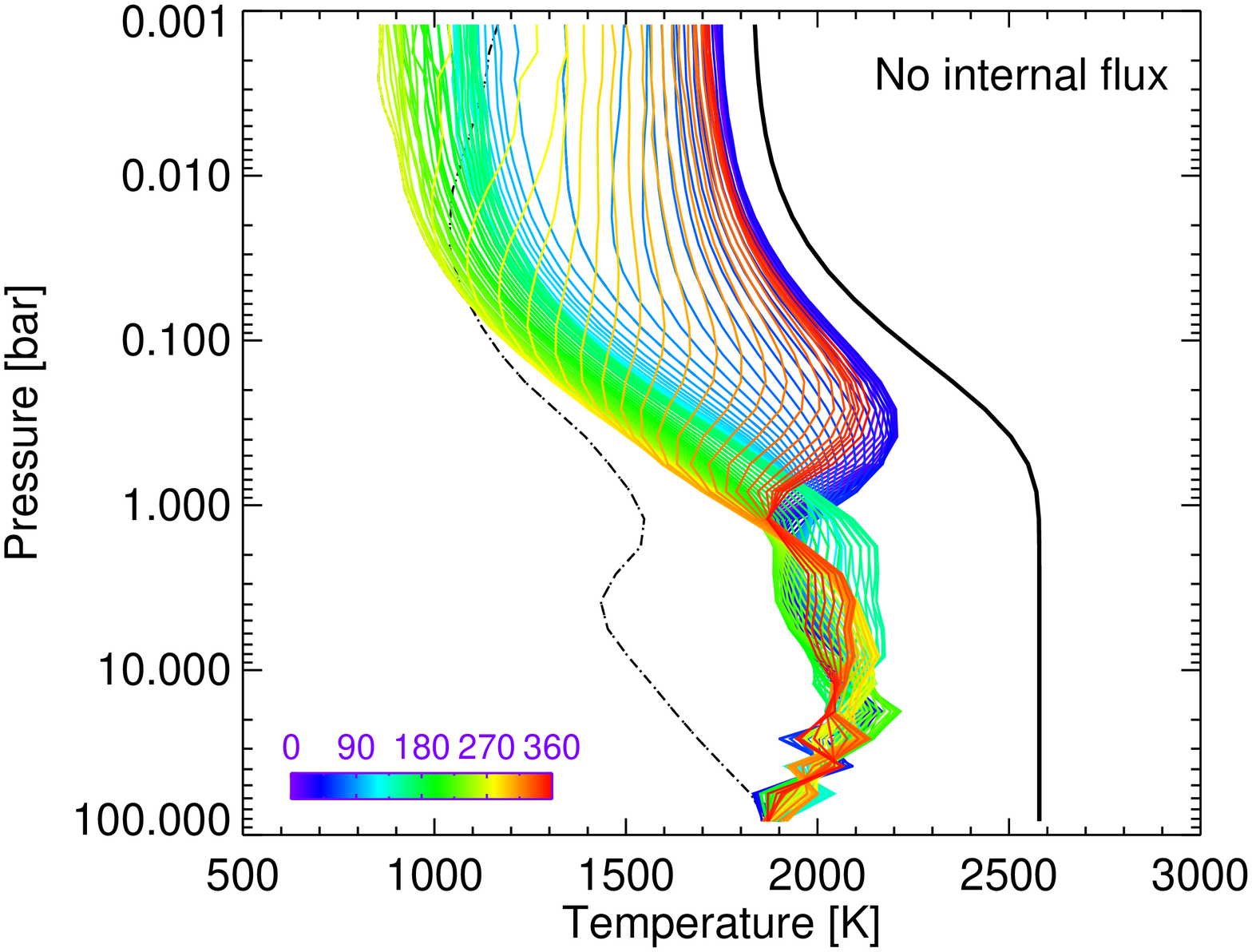}
\includegraphics[width=0.475\textwidth]{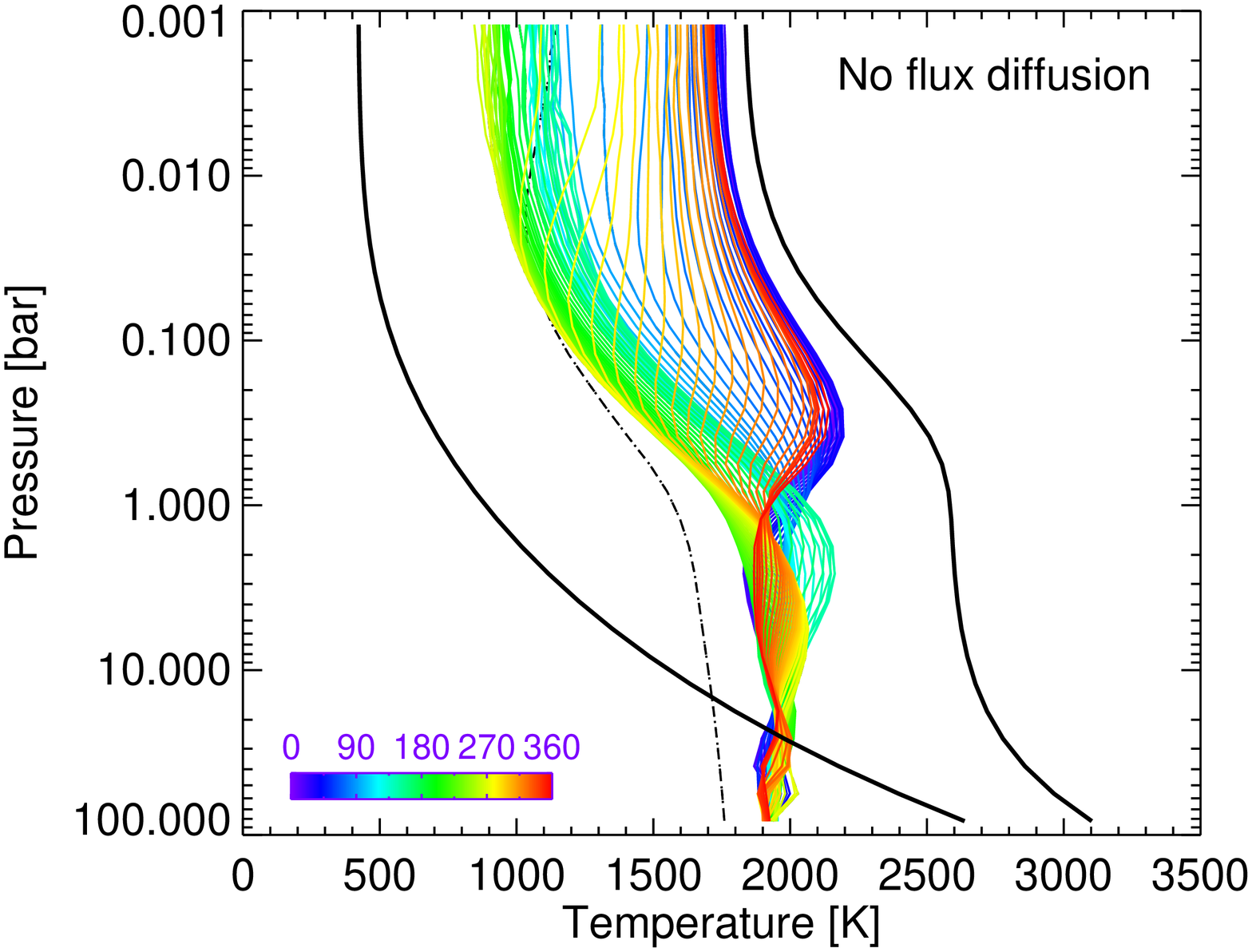}
\end{center}
\caption{Temperature-pressure profiles for two additional models: one with $\Tint=0$ K (\emph{left}) and one run without the flux diffusion scheme (\emph{right}), plotted for columns along the equator (in color) and at the poles (dotted and dashed lines), with analytic profiles for local radiative equilibrium (at the substellar point and on the nightside) shown as thick black lines.  For the model with $\Tint=0$ K, the radiative equilibrium night side temperature would be zero.  High in the atmosphere (at low pressures, above the infrared photosphere) the temperature structure is very similar between these two models and our fiducial model (compare with Figure~\ref{fig:d_tprof}).  The temperature structure of the deepest levels differs from our fiducial model, remaining isothermal rather than trending toward an interior adiabat.} \label{fig:deep}
\end{figure}

The precise value for a planet's interior heat flux is difficult to disentangle from other uncertainties in system and observed parameters.  We chose $\Tint=500$ K for our fiducial model in order to match the analytic set-up of Figure 3 from \citet{Guillot2010}, but values of \Tint~down to $\sim$100 K may be reasonable.  As a test of the influence of the choice of \Tint~on the atmospheric circulation, we ran a model with $\Tint=0$ K (and all other parameters identical to our fiducial run), with the resulting temperature structure shown in the left panel of Figure~\ref{fig:deep}.

With no heating from the interior, a night side in radiative equilibrium would have zero temperature.  In our $\Tint=0$ K model, all of the night side heating is the result of advection from the day side.  The temperatures in the upper atmosphere are still very similar to our fiducial model, a result that should not be surprising, given that the stellar heating dominates at low pressures.  The radiative fluxes for this model match those from our fiducial model (see Figure~\ref{fig:d_heating}) for pressures less than $\sim1$ bar (both globally and for different regions of the atmosphere).  Below the optical photosphere our fiducial model has upward fluxes of $\sim3500$ \Wms~(the flux corresponding to our choice of $\Tint=500$ K), while in this $\Tint=0$ K model the fluxes quickly decrease toward zero.  The heating rates for this model match our fiducial model almost identically.  The temperature structure of the deepest levels of this model remains fairly isothermal (at least along the equator).  

Similarly, the model run without the flux diffusion scheme (shown in the right panel of Figure~\ref{fig:deep}) also has an upper atmosphere whose temperature structure closely matches our fiducial model, but with an isothermal deep atmospheric structure.  Both of these tests show that the upper atmosphere (at the pressures directly probed by observations) is relatively insensitive to the conditions in the deeper atmosphere.  

However, any coupling between the atmosphere and planet interior will be sensitive to the temperature structure at deep pressures.  The isothermal behavior at deep pressures in these models means that the atmosphere has a high static stability and there is less efficient coupling between layers than in our fiducial model, where the atmosphere trends towards an adiabat.\footnote{Although we note that a radiative-convective boundary cannot exist unless the infrared opacity varies with pressure; see Appendix~\ref{sec:conv}.}  While it is not yet clear what is the best way to set the bottom boundary condition for hot Jupiter models, the temperature profiles at deep pressure will control how efficiently the atmosphere can transport energy and momentum vertically.

We note that recent work by \citet{Heng2011} uses a code very similar to the one we have developed to study a range of classes of atmospheres, including hot Jupiters.  They set their internal heat flux to zero, but choose as a bottom boundary condition to have their ``surface" layer be an idealized slab with uniform temperature and finite heat capacity, in order to simulate interaction with the planet interior.  The fluxes into and out of the bottom of their model interact with this slab by raising and lowering its temperature.  Thus, while similar, their model set-up is not identical to the one we show here.

\section{Models with radiative transfer and magnetic drag} \label{sec:drag}

Recent work suggests that hot Jupiter atmospheres should be weakly thermally ionized and---depending on the strength of the planetary magnetic field---the influence of magnetic drag on the atmospheric winds could be strong enough to significantly alter the atmospheric circulation \citep{Perna2010a}.  In the near future, observations may be able to differentiate between models with and without magnetic drag by precisely measuring the anomalous Doppler shift (due to atmospheric winds) that is present in the transmission spectra of hot Jupiters \citep{Kempton2011}.  This effect is also directly related to the structure of the planet as a whole, as ohmic dissipation of the induced currents could provide an additional heating source and possibly explain the unexpectedly large radii of some hot Jupiters \citep{Batygin2010,Batygin2011,Perna2010b}.  Simple scaling laws suggest an anti-correlation between two observable properties: the amount by which the hottest region of the atmosphere is offset from the substellar point and the degree of radius inflation \citep{Menou2011}, an issue that can be studied in more detail with numerical models.

The original models that we used to estimate the effect of magnetic drag on hot Jupiter circulation were run with the previous version of our code, in which the radiative heating was determined by a Newtonian relaxation scheme \citep{Perna2010a}.  In that work we estimated an average value for the magnetic drag timescale at each pressure level ($\tau_{\mathrm{drag}}$), for planetary magnetic field strengths of 3, 10, and 30 G.  We then simulated the effect of magnetic drag on the winds by applying a Rayleigh drag to our model, with the form: $dv/dt=-v/\tau_{\mathrm{drag}}$.  The weakest drag (for $B=3$ G) had timescales ranging from $\sim6\times10^6$ s at 1 mbar to $\sim8\times10^8$ s at 220 bar, with the medium ($B=10$ G) and strong ($B=30$ G) drag cases having timescales 10 and 100 times shorter, respectively.  Here we employ the same form for the magnetic drag, but now with our upgraded model that includes radiative transfer.  This form for the magnetic drag oversimplifies the complex physics, but gives us an initial estimate of how it may affect the circulation.  Future work will include an improved, more physically accurate, scheme for calculating the magnetic drag from local conditions, in particular allowing for horizontally non-uniform drag.

\begin{figure}[ht!]
\begin{center}
\includegraphics[width=0.325\textwidth]{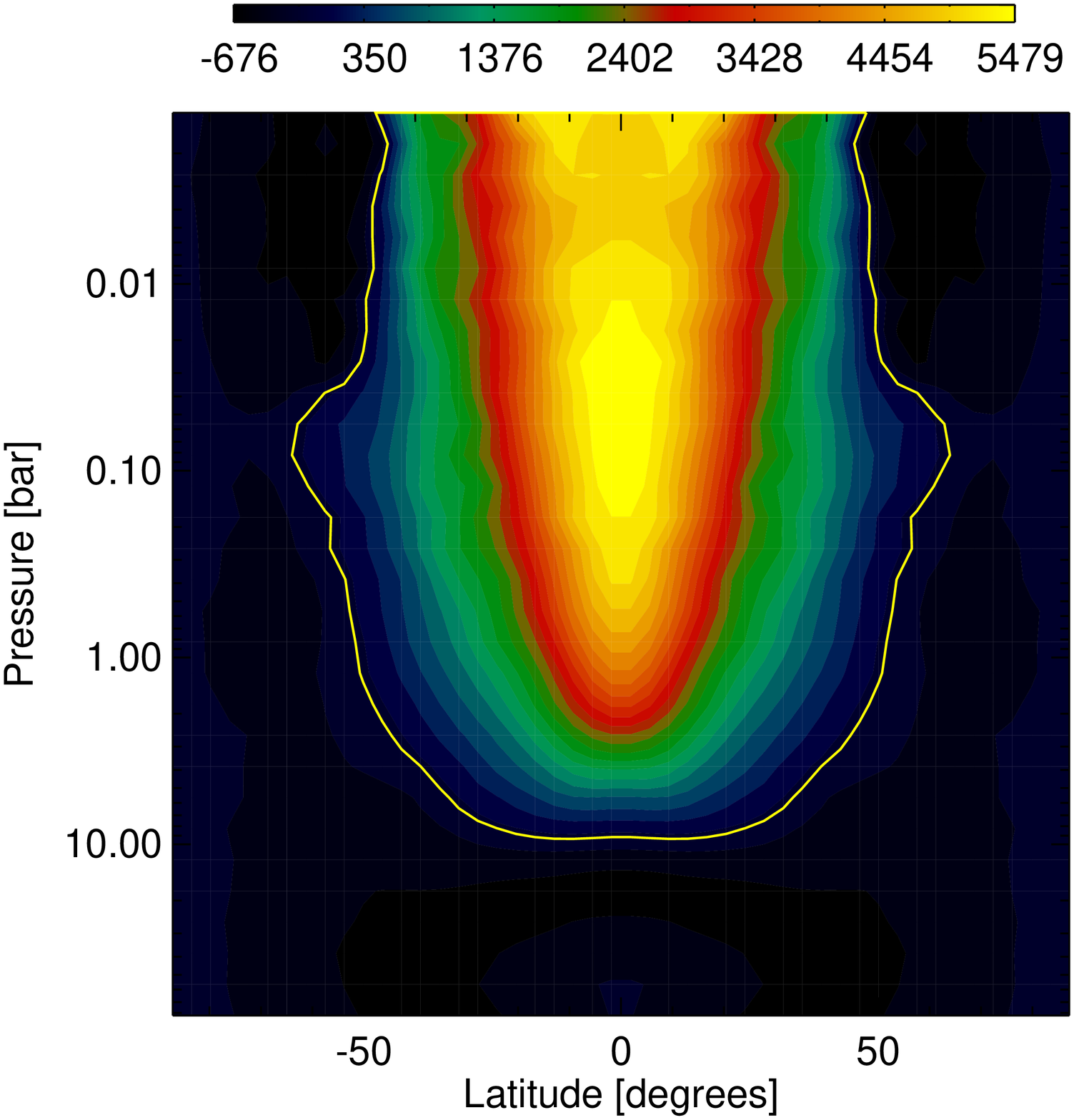}
\includegraphics[width=0.325\textwidth]{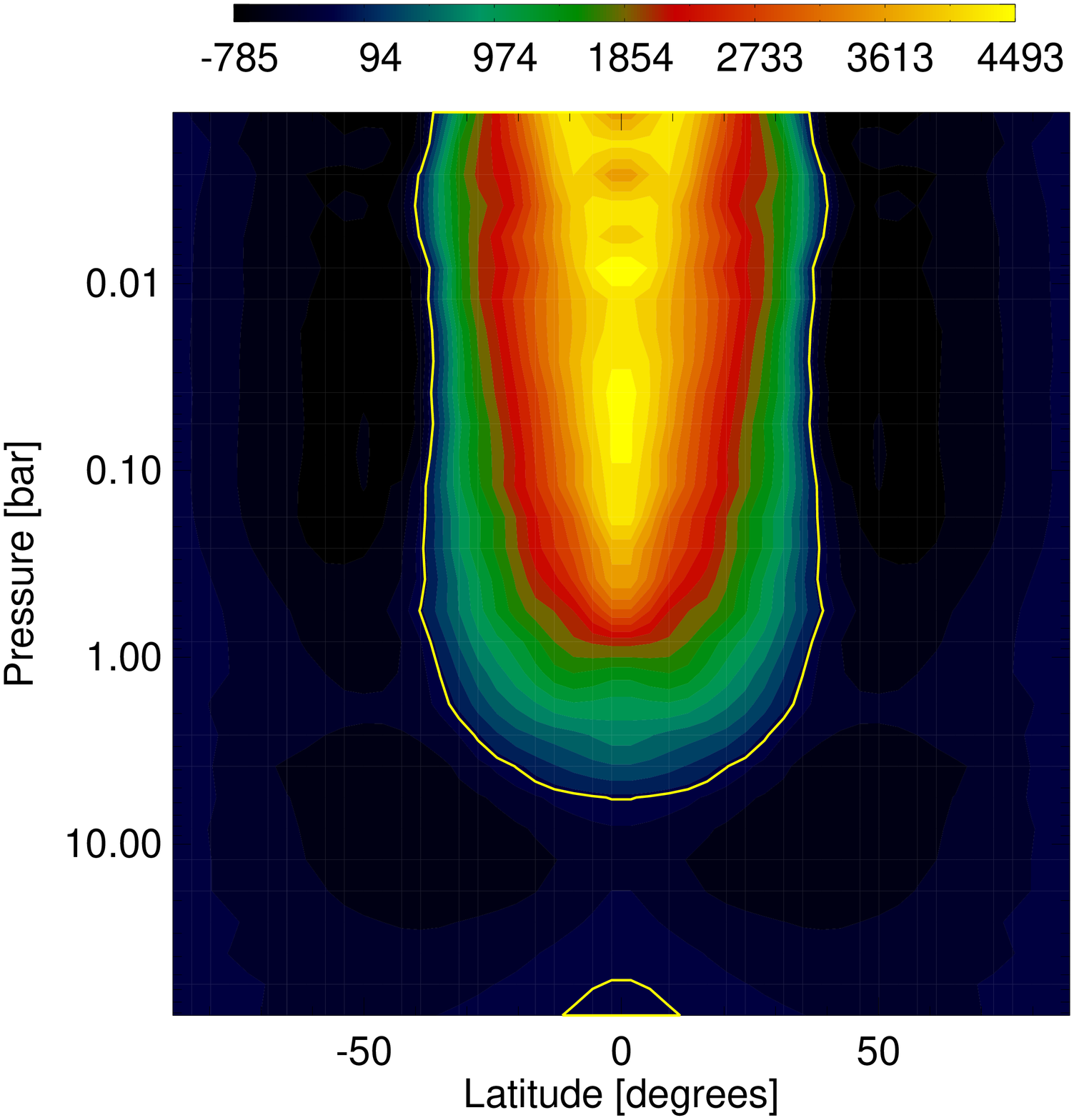}
\includegraphics[width=0.325\textwidth]{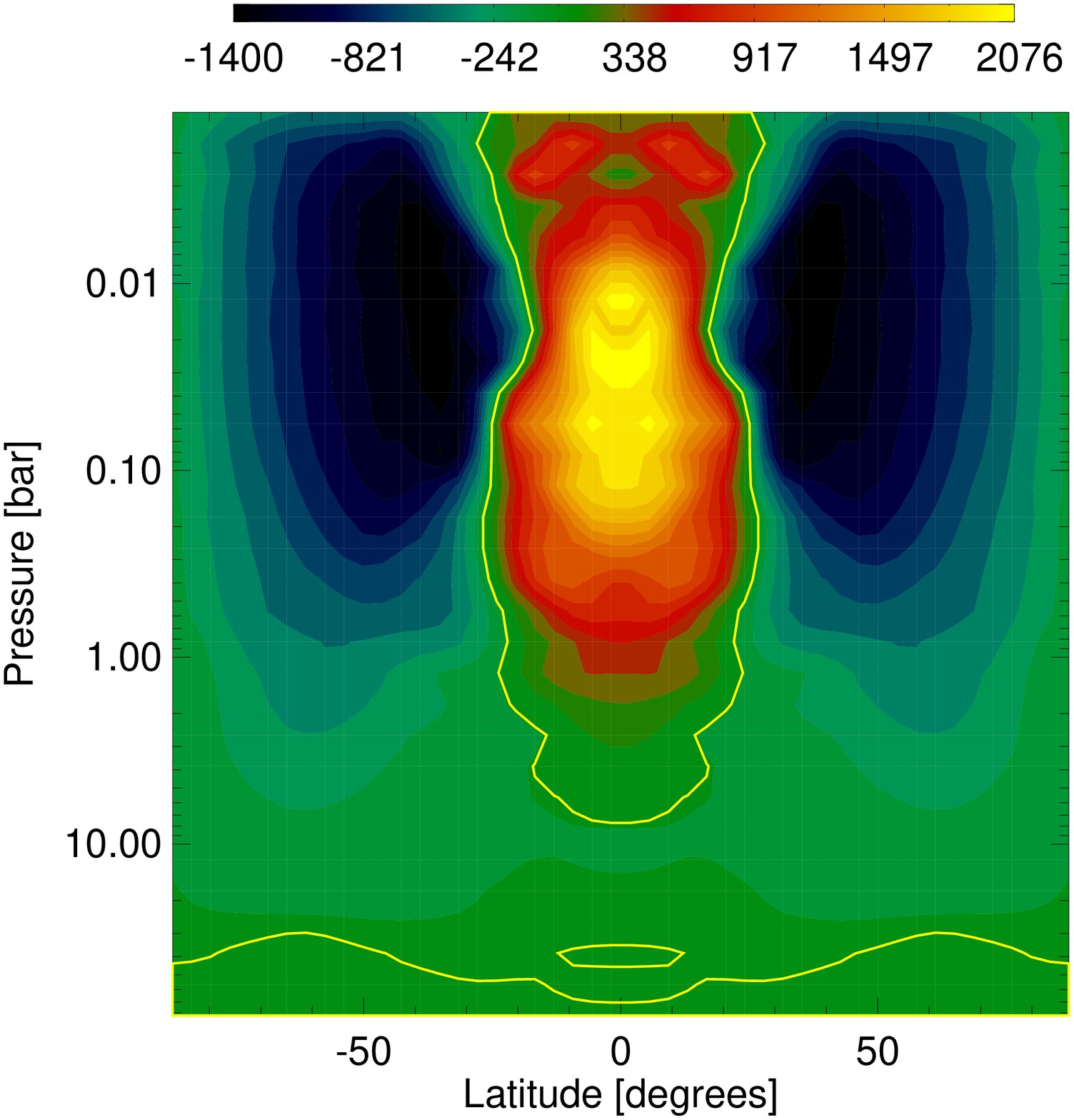}
\end{center}
\caption{Plots showing the zonal average of the zonal (east-west) wind, as a function of latitude and pressure, for models with weak (\emph{left}), medium (\emph{middle}), and strong (\emph{right}) drag.  The yellow line separates eastward (positive) from westward (negative) flow.  In comparison with the drag-free fiducial model (Figure~\ref{fig:d_uz}), we see that the effect of increasing drag is to slow wind speeds and to restrict the eastward equatorial jet to lower pressures.  The dynamical structure of the atmosphere begins to change at the strongest level of drag, and is no longer dominated by eastward flow.} \label{fig:drag_uz}
\end{figure}

In Figure~\ref{fig:drag_uz} we show the zonal wind profiles for models with weak, medium, and strong drag.  By comparison with our fiducial, drag-free model (Figure~\ref{fig:d_uz}) we can see that increased drag has three effects on the flow: (1) slower wind speeds, (2) less of a disparity in strength between eastward and westward winds, and (3) the confinement of the super-rotating equatorial jet to higher pressures.  This is similar to our results in Figure~3 of \citet{Perna2010a}.  Note that the model with the strongest amount of drag still has supersonic winds in the upper atmosphere ($\sim7$ \kms), even though the zonally averaged wind speeds in Figure~\ref{fig:drag_uz} are subsonic ($c_s \approx 2-3$ \kms); this is because there is strong substellar-to-antistellar flow and the eastward and westward winds cancel each other in the zonal average.

\begin{figure}[ht!]
\begin{center}
\includegraphics[width=0.95\textwidth]{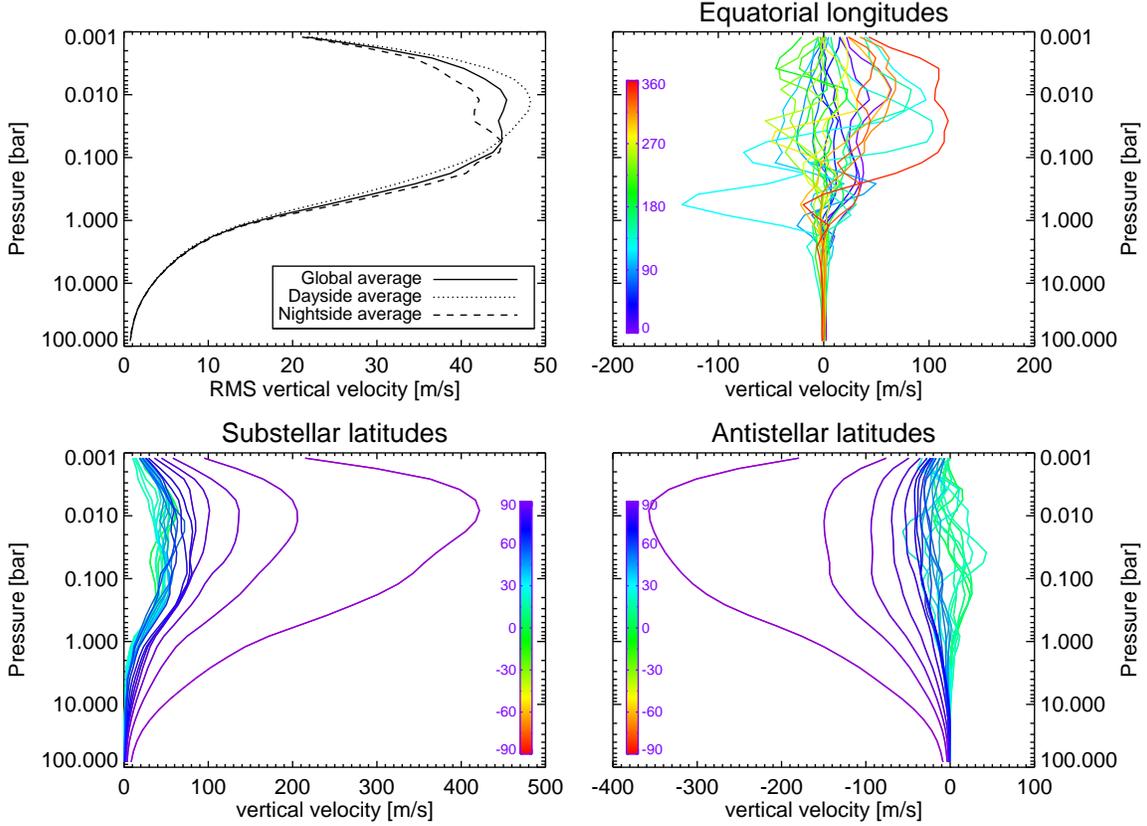}
\end{center}
\caption{Vertical velocities in the model with the strongest drag.  For a description of each plot, see Figure~\ref{fig:d_vert}, the same set of plots for our fiducial (drag-free) model.  Although horizontal wind speeds are reduced by the drag, the vertical velocities in this model are generally increased, relative to the fiducial model.} \label{fig:drag_vert}
\end{figure}

The structure of the horizontal flow changes with increasing drag strength, as does the structure of the vertical flow.  In Figure~\ref{fig:drag_vert} we show profiles of the vertical winds throughout the model with strong drag.  Comparing with the fiducial model (Figure~\ref{fig:d_vert}) we see that two of the main trends remain: vertical motion is generally stronger at lower pressure and is strongest at the poles.  Some of the details of the flow do change; in particular, at deep pressures we see slower winds and less of a difference between average speeds on the day and night side.  Finally, although the horizontal wind speeds have been reduced by the strong drag, the drag does not act on vertical motion and the vertical winds at low pressure are slightly stronger in this model than in the fiducial one.  This could lead to a slight change in the ability of the atmosphere to vertically mix chemical species.

\begin{figure}[ht!]
\begin{center}
\includegraphics[width=0.45\textwidth]{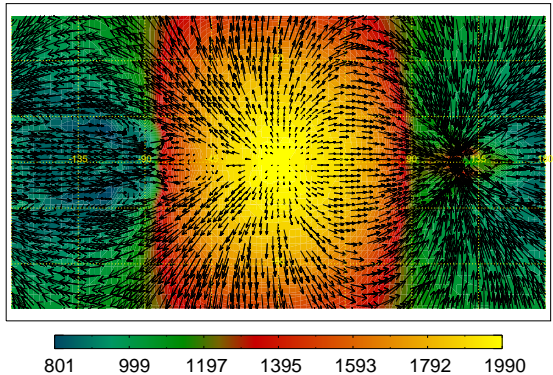}
\end{center}
\caption{The infrared photosphere of the planet for our model with strong magnetic drag, shown in cylindrical projection with the substellar point at the center of the plot.  The color gives the temperature [in K] and the arrows show the wind vectors.  The maximum wind speed at this level is 6 \kms.  While there is an eastward equatorial jet at this level, it is prevented from completely circling the globe by the strong substellar-to-antistellar flow.  This results in a important difference from our fiducial, drag-free model (Figure~\ref{fig:d_phot}): there is no significant advection of the hottest region of the atmosphere eastward of the substellar point.} \label{fig:drag_phot}
\end{figure}

In the model with strong drag, the circulation pattern is transitioning away from the behavior seen in the low- and no-drag models.  At high altitude the substellar-to-antistellar flow dominates, with the eastward equatorial jet confined to the night side.  In fact, even though the zonal average of equatorial winds is eastward for much of the atmosphere, there are no pressure levels at which eastward equatorial flow is able to completely circle the globe.  In Figure~\ref{fig:drag_phot} we plot the temperature and wind structure at the photosphere, for the strong drag model.  The slower wind speeds and the transition to a primarily substellar-to-antistellar flow structure results in a hot day/cold night temperature structure.  In the strong drag model the hottest region of the atmosphere remains at the substellar point, rather than being advected eastward, as is the case for our fiducial model (compare to Figure~\ref{fig:d_phot}).

\begin{figure}[ht!]
\begin{center}
\includegraphics[width=0.42\textwidth]{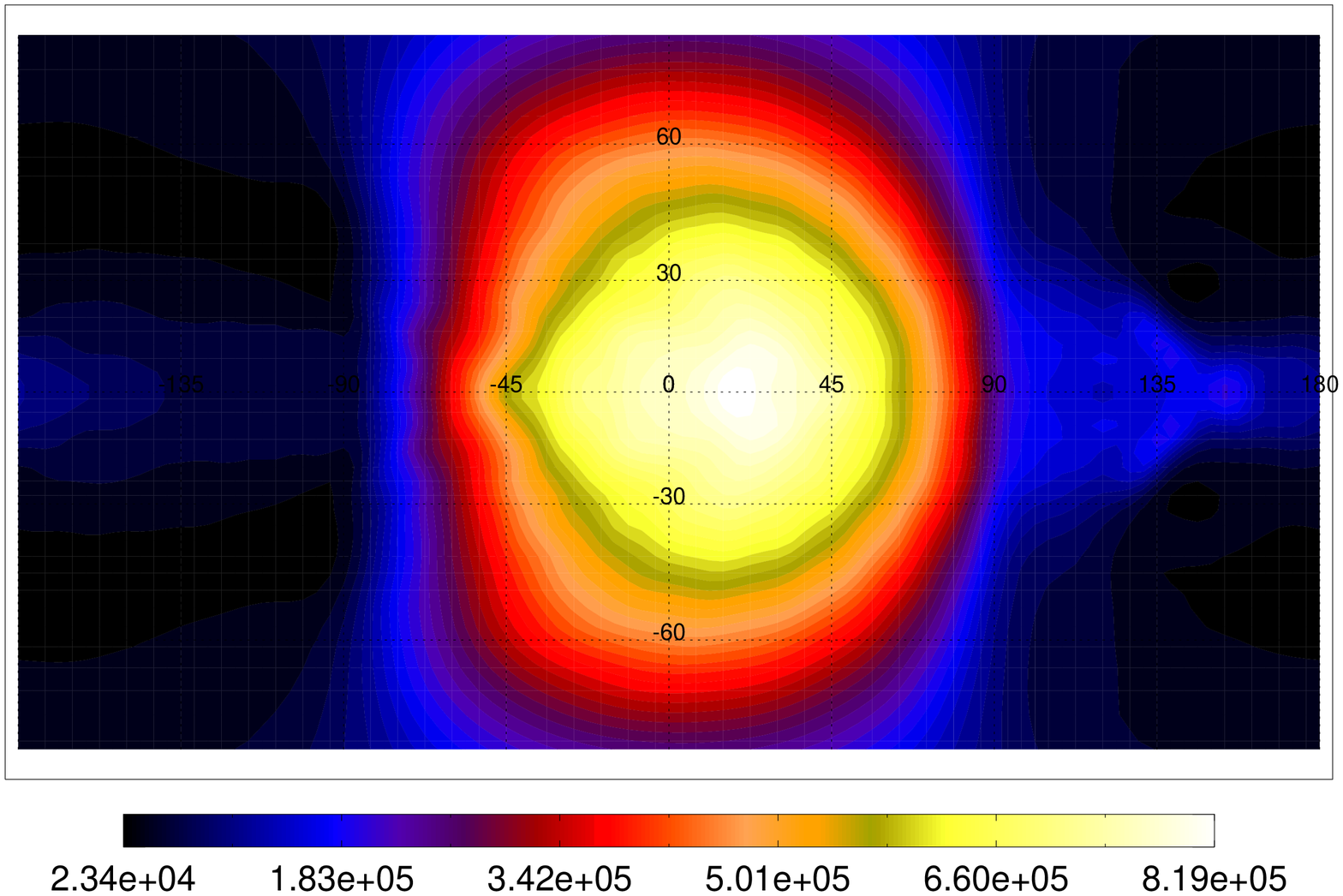} \\
\includegraphics[width=0.42\textwidth]{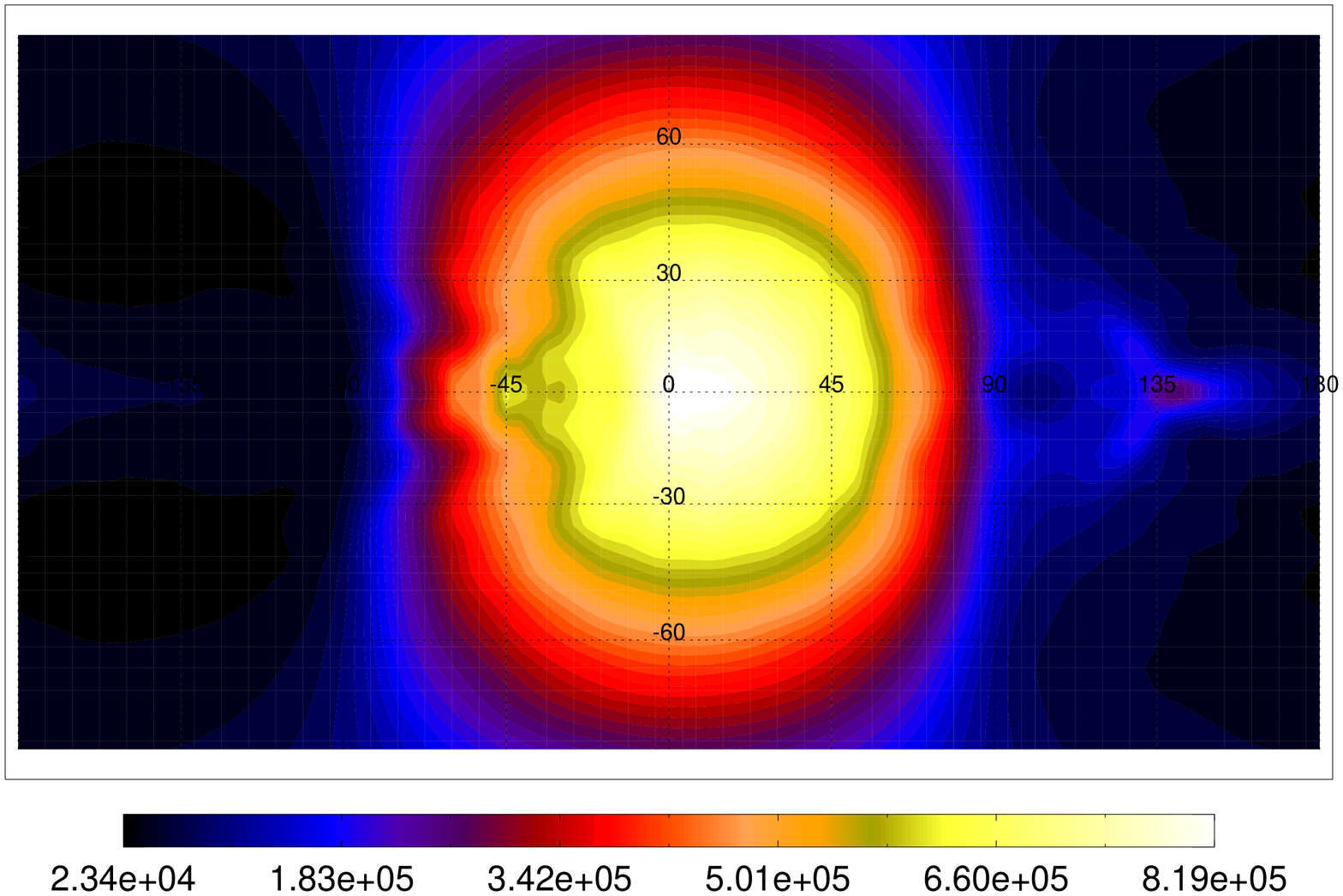} \\
\includegraphics[width=0.42\textwidth]{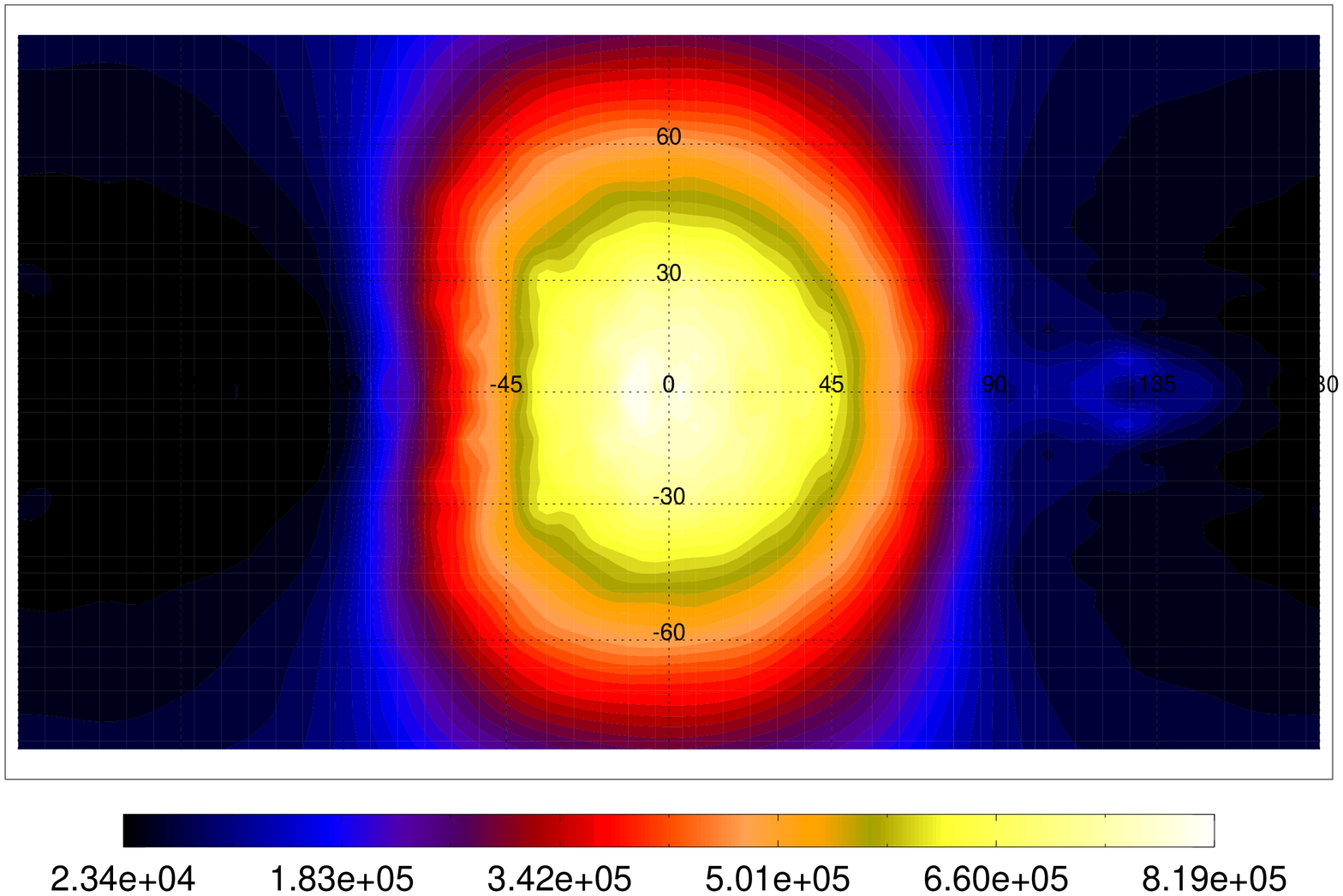}
\end{center}
\caption{Cylindrical maps of the infrared flux [in \Wms] emitted from the top boundary of the models with weak (\emph{left}), medium (\emph{middle}), and strong (\emph{right}) drag (compare to Figure~\ref{fig:d_phot} for the drag-free model).  As the amount of drag is increased, the brightest region of the atmosphere remains closer to the substellar point (rather than being advected eastward) and the night side is dimmer, due to the winds being less efficient at decreasing the day-night temperature contrast.}  \label{fig:drag_folr}
\end{figure}

\begin{figure}[ht!]
\begin{center}
\includegraphics[width=0.8\textwidth]{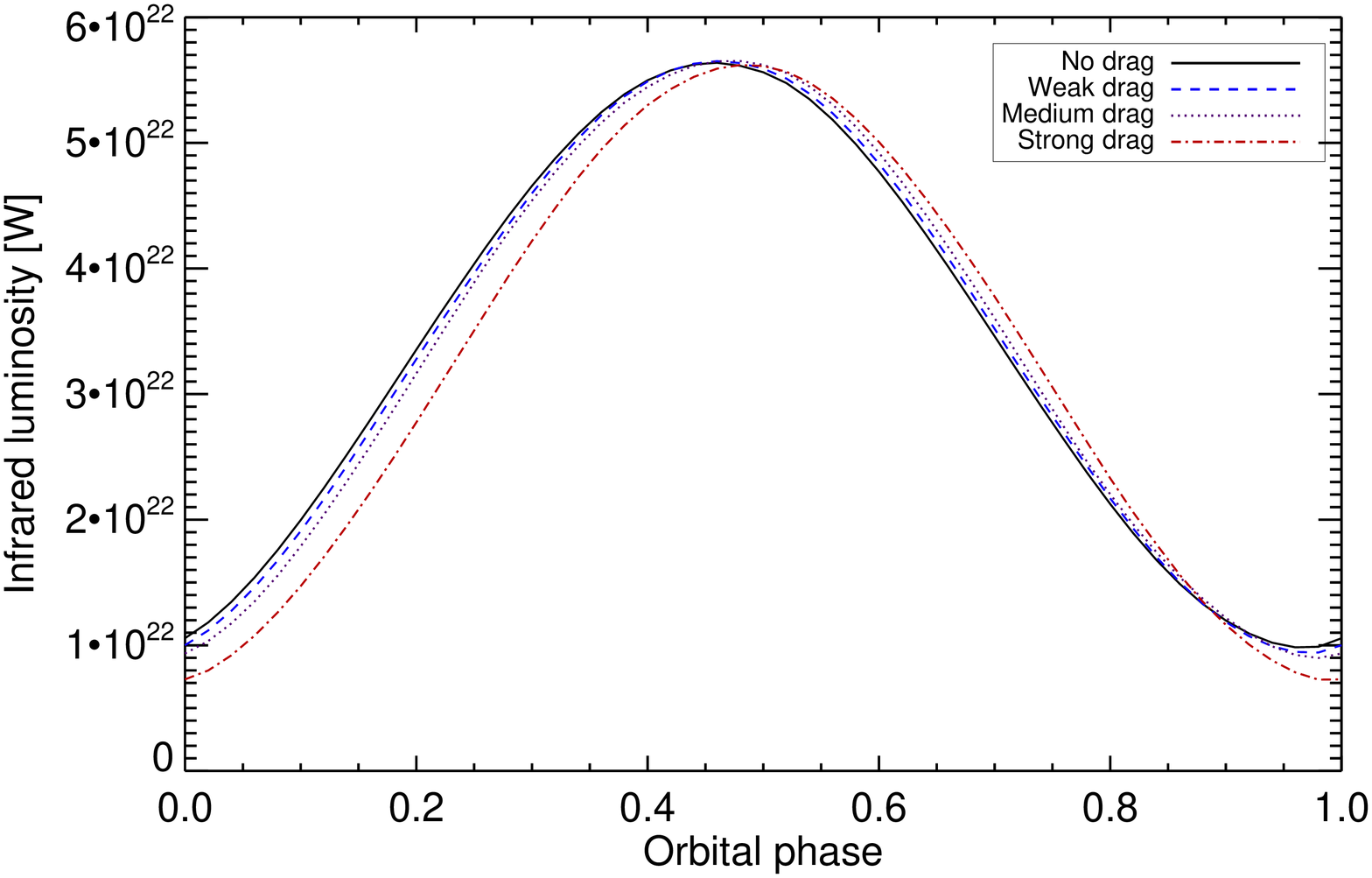}
\end{center}
\caption{Emitted infrared phase curves for our fiducial model without drag (\emph{black}), and the models with weak drag (\emph{dashed blue}), medium drag (\emph{dotted purple}), and strong drag (\emph{dash-dotted red}).  Secondary eclipse (not shown) would occur at an orbital phase of 0.5.  The maximum luminosity from each model occurs at an orbital phase of 0.467, 0.469, 0.481, and 0.494, which through a simple conversion would correspond to the hottest region of the atmosphere at a planet longitude of 12\degrees, 11\degrees, 7\degrees, and 2\degrees, respectively.} \label{fig:pc}
\end{figure}

As a result of the change in the circulation pattern, we find as expected a correlation between increased drag strength and a decreased shift of the hotspot away from the substellar point; in Figure~\ref{fig:drag_folr} we show maps of the infrared flux emitted by the planet for the weak-, medium-, and strong-drag models.  By integrating over the visible portion of the planet, for viewing angles along the equator, we calculated the infrared luminosity that would be observed throughout the planet's orbit, for our fiducial and the three drag models (Figure~\ref{fig:pc}).  We used a single model snapshot, instead of snapshots of the planet throughout an orbit, but the level of atmospheric variability is low enough in all models that the shape of the light curve should be unaffected.  

We find very little difference between the phase curves for the fiducial, weak-, and medium-drag models; however, the peak in flux from the strong-drag model is well aligned with an orbital phase of 0.5 (when the substellar point is directed toward the observer).  The peak in luminosity occurs at orbital phases ranging from 0.467 for the drag-free model to 0.494 for the strong-drag model.  It may be challenging to observationally distinguish between these models with current instruments.  The best hot Jupiter thermal phase curves so far can only constrain the phase of peak flux to $\pm0.02$ \citep{Knutson2007,Knutson2009}.  

The phase curve for the strong-drag model is also more sharply peaked over the day side and has a dimmer night side than the other three models, leading to a larger flux contrast between the day and night sides.  The fiducial model has a minimum flux that is 17.4\% of the maximum, while this ratio is 12.9\% for the strong-drag model.  The most sensitive phase curve observations can measure this ratio to within $\pm$ 3\% - 7\% \citep{Knutson2007,Knutson2009}, meaning that this measurement may also be challenging, if one is to discriminate between these models.

In order to compare the observable properties of radius inflation and hotspot offset \citep{Menou2011}, we must estimate the ohmic heating that would be produced in these magnetic drag models.  Although we do not explicitly calculate the heating from ohmic dissipation in these models, it should be equal to the amount of kinetic energy dissipated by magnetic drag, which we can calculate from the kinetic energy fields and prescribed drag timescales.  We report this value for each of our models in Table~\ref{tab:energy}.  As a percentage of the stellar heating input to our models, we find that the ohmic heating in our weak-, medium-, and strong-drag models is at efficiencies of 0.6\%, 3\%, and 60\%.  The amount of extra heating necessary to significantly inflate a hot Jupiter's radius depends, in part, on the pressures at which it is deposited \citep[e.g.,][]{Guillot2002}.  Ohmic heating is a non-local process, meaning that the dissipation of induced currents does not necessarily happen in the same location as where magnetic drag brakes the winds, and in this paper we do not estimate where the heating may occur; however, our previous work indicated that enough of the heating may take place deep enough in the atmosphere to contribute to radius inflation \citep{Perna2010b}.  Although it is not clear if the weak-drag model would produce enough ohmic heating be significant, the 3\% efficiency found in our medium-drag model should be able to inflate the planetary radius, while the 60\% efficiency of our strong-drag model is so high that this heating may be able to completely evaporate the planet \citep[according to the models of][]{Batygin2011}.  It is important to note that our strong drag model is not self-consistent, in that the ohmic heating is not explicitly included and it would likely have a large effect on the circulation pattern and therefore the amount of kinetic energy dissipated by drag.  A better analysis of the correlation between the eastward shift of the hottest region of the atmosphere and the amount of radius inflation must wait for an improved modeling scheme which accounts for magnetic drag and its associated ohmic heating self-consistently.

Finally, we calculate the rate of numerical kinetic energy loss in each of our drag models.  For our fiducial model the rate of kinetic energy loss was found by comparing the non-zero global net radiative heating to the rate of energy input to the atmosphere, assuming that the net heating was balancing the net numerical loss.  The magnetic drag is an explicit sink of kinetic energy and we subtract this value from the net radiative heating, assuming that the remainder is balancing numerical dissipation.  As we report in Table~\ref{tab:energy}, the amount of numerical loss is similar between our fiducial, weak-, and medium-drag models, but is zero (to within 1\%) in our strong-drag model.  An exact identification of the source of numerical loss in our models (e.g., whether it is localized or distributed uniformly) is beyond the scope of this paper; however, the lack of any significant loss in our strong-drag model seems to indicate that the numerical loss is related to the much stronger wind speeds in our other models.  It may be that we can substantially reduce numerical loss by correctly including all sources of drag in our models, be it magnetic or related to subgrid processes \citep[see, e.g.,][]{Li2010}.

\begin{deluxetable}{lcccc}
\tablewidth{0pt}
\tablecaption{Heating and dissipation rates}
\tablehead{
\colhead{Model} &  \colhead{Rate of explicit kinetic}  & \colhead{Globally integrated} & \colhead{Rate of numerical}  \\
\colhead{}  & \colhead{energy dissipation} & \colhead{net heating rate} & \colhead{kinetic energy loss} 
}
\startdata
Fiducial, drag-free 	& 0					&  $3.9\times 10^{21}$ W		& 12\%  \\
Weak drag		& $2.0\times 10^{20}$ W &   $5.6\times 10^{21}$ W		& 16\%  \\
Medium drag		& $1.1\times 10^{21}$ W &   $6.2\times 10^{21}$ W		& 15\%  \\
Strong drag		& $2.0\times 10^{22}$ W &   $2.0\times 10^{22}$ W		& 0\% \\
\enddata
\label{tab:energy}
\tablecomments{The explicit dissipation due to magnetic drag ($d_{\mathrm{drag}}$) is calculated from the kinetic energy field and the drag timescale used the model.  The global net heating rate ($q_{\mathrm{net}}$) is a total of all radiative heating and cooling over the entire atmosphere.  The heating input to the atmosphere ($q_{\mathrm{input}}$) is the total stellar radiation incident on the top boundary ($3.3\times10^{22}$ W), minus the cooling flux coming up through the bottom boundary ($4.4\times10^{20}$ W).  From the assumption that the nonzero net heating is balancing the dissipation rate (from drag plus numerical effects), we calculate the rate of numerical loss as $=(q_{\mathrm{net}} - d_{\mathrm{drag}})/q_{\mathrm{input}}$.  See the text for discussion.}
\end{deluxetable}
\clearpage

\section{Summary} \label{sec:conc}

We have presented a new radiative transfer scheme for our atmospheric circulation code.  It divides all radiation into optical and infrared wavelengths.  The optical flux is incident on the top boundary and its attenuation is controlled by a constant optical absorption coefficient.  The infrared flux is absorbed and emitted at each level, as governed by an infrared absorption coefficient that goes as a powerlaw with pressure (and can be constant).  Below the infrared photosphere the infrared fluxes are calculated using the flux-limited diffusion approximation, which we found was necessary in order to reproduce analytic temperature profiles correctly.

Using this new code, we present a fiducial model for a generic hot Jupiter and find that the temperature and wind structures agree well with previously published models.  We also show vertical velocities, radiative fluxes, and heating/cooling rates as a function of pressure and region of the atmosphere.  Our radiative scheme allows us to self-consistently predict the infrared flux emitted by the planet and we map this as a function of location on the globe, reproducing the standard shift of the brightest region eastward of the substellar point.

We show temperature profiles for models that: 1) did not use the flux diffusion scheme, or 2) assumed zero flux from the interior.  In both cases the deepest pressure levels are mostly isothermal, instead of increasing in temperature toward the inner convective zone, significantly changing the ability of the atmosphere to exchange energy and momentum with the interior.  However, the upper atmosphere (including and above the infrared photosphere) are relatively insensitive to the behavior in the deepest levels and have similar observable properties to our fiducial model.

We present models that use our new radiative transfer scheme in combination with a simplified form of magnetic drag, as an improvement on the models presented in \citet{Perna2010a}.  As in those earlier models, we find that as the drag strength is increased, the structure of the atmosphere changes; the strongest level of drag is able to significantly alter the atmosphere from its drag-free state.  As a new result, we are able to measure the amount by which the brightest region of the atmosphere is offset from the substellar point, as a function of the strength of the magnetic drag used in each model.  The weak- and medium-drag models have orbital thermal phase curves similar to the drag-free model.  The winds are slowed enough in the strong-drag model that the brightest region of the atmosphere is closely aligned with the substellar point and, related to this effect, the day-night flux amplitude in the strong-drag model is slightly larger than for the drag-free model.  We estimate the amount of ohmic heating that would be produced for each of our models and find that the medium-drag model should have a significantly inflated planetary radius, while the amount of heating in the strong-drag model is so high that it could perhaps lead to planet evaporation.  A better analysis of magnetic effects on hot Jupiter atmospheres must wait for improvements in our code, however, using a more realistic form for the drag and explicitly including the effect of ohmic heating.

Finally, we estimate the rate of numerical loss of kinetic energy in each of our models and find it to be at the level of $\sim$15\%, except for our model with strong magnetic drag, which we calculate to have losses consistent with zero at the percent level.  Our strong-drag model has much slower winds than the other models presented here and we speculate that strong numerical loss is directly related to those high speed winds. 

\acknowledgements

This work was performed in part under contract with the California Institute of Technology (Caltech) funded by NASA through the Sagan Fellowship Program.  KM was supported by NASA grant PATM NNX11AD65G.

\appendix

\section{Analytic solutions for a non-constant infrared opacity} \label{sec:tpprofiles}

Equation 27 of \citet{Guillot2010} gives the temperature profile for an atmosphere with constant visible and infrared opacities:
\begin{equation}
T^4 = \frac{3\Tint^4}{4} \left[ \frac{2}{3}+\tau \right] + \frac{3 \Tirr^4}{4} \mu_\star \left[ \frac{2}{3} + \frac{\mu_\star}{\gamma} + \left( \frac{\gamma}{3\mu_\star}-\frac{\mu_\star}{\gamma}\right) e^{-\gamma \tau/\mu_\star} \right].
\end{equation}
\noindent  The atmosphere is heated from below by an interior heat flux, $F_{\mathrm{int}}=\sigma_{\mathrm{SB}} \Tint^4$.  The stellar flux incident on the top of the atmosphere has a strength of $\sigma_{\mathrm{SB}} \Tirr^4$ at the substellar point and decreases as $\mu_\star$, the cosine of the angle from the substellar point.  The irradiation temperature is: $\Tirr=T_\star (R_\star/a)^{1/2}$, where $a$ is the planet-star distance.  For given heating strengths, the temperature profile depends on the ratio of visible and infrared opacities, $\gamma=\kvis/\kir$, and is a function of the (infrared) optical depth $\tau=(\kir/g)P$.

If we let the infrared opacity vary as a powerlaw (Equation~\ref{eqn:kir}), then $\gamma$ is now a function of pressure and the optical depth is no longer linear with pressure: $\tau=(\kappa_{\mathrm{IR},0}/g)P^{\alpha+1}$.  In this case, the exponential term in Equation 21 of \citet{Guillot2010} can no longer be solved analytically and an altered derivation results in a new form for the temperature profile:
\begin{equation}
T^4 = \frac{3\Tint^4}{4} \left[ \frac{2}{3}+ \frac{\tau}{\alpha+1} \right] + \frac{3 \Tirr^4}{4} \mu_\star \left[ \frac{2}{3} + \frac{\gamma}{3\mu_\star} e^{-\gamma \tau/\mu_\star} + \frac{\kappa_{\mathrm{IR},0}}{g} \int_0^P \left(\frac{P'}{P_{\mathrm{ref}}}\right)^{\alpha} e^{-\gamma \tau/\mu_{\star}} dP' \right].  \label{eqn:tprof}
\end{equation}
\noindent  Note that $\gamma \tau = (\kvis/g) P$ is the optical depth for visible wavelengths.  Instead of solving for the temperature profile as a function of $\mu_\star$, we could use the averaging factor $f$ \citep[see Equation 29 in][]{Guillot2010}, which gives substellar, dayside-averaged, or globally-averaged profiles.  In that case our derivation of Equation~\ref{eqn:tprof} agrees with the derivation of Equation 31 in \citet{Heng2011}, which likewise allows for a non-constant infrared opacity.

\section{Opacity conditions necessary for a convective region to exist} \label{sec:conv}

An atmosphere subject to external heating will be statically stable at low pressures and will transition to an interior convective zone at high pressures.  In order for a convective temperature-pressure profile to exist within our modeling scheme, the optical and infrared opacities cannot both be constant.  Here we demonstrate this analytically and derive conditions for the infrared opacity function such that the atmosphere can be convective.

If both the infrared and optical absorption coefficients are constant with pressure, the temperature-pressure profiles will never include a radiative-convective boundary, regardless of how deep the bottom boundary is set or how high of an internal heat flux is used.  From Equation 27 of \citet{Guillot2010} we can calculate that the night side temperature-pressure profile\footnote{At depth the day and night side profiles will match.} for an atmosphere with constant absorption coefficients will follow: 
\begin{equation}
\frac{ d \ln T}{d \ln P} = \frac{\tau}{4(2/3 + \tau)},
\end{equation}
\noindent and as the optical thickness goes to infinity, $d (\ln T)/ d (\ln P)$ reaches a maximum value of 0.25.  Convection occurs when $d (\ln T)/ d (\ln P) \geq R/c_p$; in the case of a diatomic gas $R/c_p=0.286$ and an atmosphere with constant absorption coefficients will never be convective.

For an atmosphere in which the infrared absorption coefficient scales exponentially with pressure (as per Equation~\ref{eqn:kir}), the formalism of \citet{Guillot2010} can be expanded (see Equation \ref{eqn:tprof}) to find that the night side profile will follow:
\begin{equation}
\frac{ d \ln T}{d \ln P} = \frac{\tau}{4(2/3 + \tau/[\alpha+1])}
\end{equation}
\noindent where the optical depth is no longer linear with pressure: $\tau=(k_{\mathrm{IR,0}}/g)(P/P_{\mathrm{ref}})^\alpha$.  As the optical depth goes to infinity, $d (\ln T)/ d (\ln P)$ reaches a maximum value of $(\alpha+1)/4$.  The atmosphere will be convective at depth when: $(\alpha+1)/4 \geq R/c_p$; for the case of a diatomic gas this requires $\alpha \geq 1/7$.

Our code transitions to using fluxes from the diffusion approximation in the deep, optically thick atmosphere.  \citet{Arras2006} solve for analytic pressure-temperature profiles deep in a gas giant atmosphere, assuming flux-limited diffusion and an absorption coefficient that scales as a powerlaw with pressure (and temperature).  Their requirement for a convective zone to exist ($\nabla_\infty \geq \nabla_{\mathrm{ad}}$), when converted into our notation, is: $(\alpha+1)/4 \geq R/c_p$.  This is consistent with the result above and sets a robust requirement for our models.

\end{document}